\documentclass[lettersize, journal]{IEEEtran}
\IEEEoverridecommandlockouts

\usepackage{setspace}
\usepackage{cite}
\usepackage{amsmath,amssymb,amsfonts}
\usepackage{algorithm}
\usepackage{algorithmic}
\usepackage{graphicx}
\usepackage{comment}
\usepackage{textcomp}
\usepackage{subcaption}

\usepackage{xcolor}
\usepackage{stfloats}
\def\BibTeX{{\rm B\kern-.05em{\sc i\kern-.025em b}\kern-.08em
    T\kern-.1667em\lower.7ex\hbox{E}\kern-.125emX}}
\usepackage{hyperref}

\begin{document}

\title{{Alternating Optimization for Joint Resource Allocation in Full-Duplex Multi-Sector Fluid Antenna-Enabled Near-Field Systems}\\
}

\author{Jingxuan Zhou,~\IEEEmembership{Student Member,~IEEE,}
        Yinchao Yang,~\IEEEmembership{Student Member,~IEEE,}
        Zhaohui Yang,~\IEEEmembership{Member,~IEEE,}
        A. Hamid Aghvami,~\IEEEmembership{Life Fellow,~IEEE,}
        and~Mohammad~Shikh-Bahaei,~\IEEEmembership{Senior Member,~IEEE}
\thanks{This work is supported in part by Supported by National Key R\&D Program of China (Grant No. 2023YFB2904804), Zhejiang Key R\&D Program under Grant 2023C01021, Young Elite Scientists Sponsorship Program by China Association for Science and Technology under Grant 2023QNRC001, the Fundamental Research Funds for the Central Universities under Grant 226-2024-00069.}
\thanks{J. Zhou, Y. Yang, A. H. Aghvami, and M. Shikh-Bahaei are with the Department of Engineering, King's College London, Strand, London WC2R 2LS, UK (E-mail: jingxuan.zhou@kcl.ac.uk, yinchao.yang@kcl.ac.uk,  hamid.aghvami@kcl.ac.uk, m.sbahaei@kcl.ac.uk)}
\thanks{Z. Yang is with the College of Information Science and Electronic Engineering, Zhejiang University, Hangzhou 310027, China, International Joint Innovation Center, Zhejiang University, Haining 314400, China, and with Zhejiang Provincial Key Laboratory of Info. Proc., Commun. \& Netw. (IPCAN), Hangzhou 310007, China (E-mail: yang\_zhaohui@zju.edu.cn)}
}

\maketitle

\begin{abstract}
This paper proposes a full-duplex fluid-antenna near-field system (FD-FANS) with a multi-sector antenna array that jointly exploits resource allocation, antenna mobility, and group-based transmitting (TX) and receiving (RX) partitioning. A spherical-wave uplink/downlink channel is established that accounts for residual self-interference (SI), wireless energy transfer (WET), and geometric constraints on antenna motion. Within the FD-FANS framework, an efficient protocol is devised to enable simultaneous downlink energy transmission (DET) and uplink data transmission (UDT) at the base station (BS). Furthermore, we formulate, for both perfect and imperfect SI cancellation (SIC), a weighted sum-rate (WSR) maximization problem over time–power allocation, antenna positions, and binary group selection, under practical average and peak power limits, per-antenna box constraints, minimum spacing, and a half-TX/half-RX balance. To tackle the resulting non-convex mixed-integer design, we develop an efficient alternating-optimization (AO) framework based on majorization–minimization / successive convex approximation (MM/SCA). The proposed algorithm monotonically improves the objective and converges to a stationary solution of a continuous relaxation. Simulation results demonstrate that the proposed scheme achieves consistent performance gains over several benchmark designs, including half-duplex FANS (HD-FANS), FD fixed-position antenna near-field system (FD-FPANS), non-grouped FD-FANS, and far-field counterparts, in terms of average sum-rate (ASR), energy efficiency (EE), and user fairness, while exhibiting robustness to residual SI and channel uncertainty.
\end{abstract}


\section{Introduction} \label{s1}

The upcoming sixth-generation (6G) wireless networks are envisioned to meet unprecedented performance metrics, including terabit per second (Tb/s) data rates, millisecond-level latency, ultra-reliable connectivity, and enhanced energy efficiency \cite{wang2023road,2024arXiv241202538Y, bobarshad2009m, bobarshad2010low, nehra2010spectral, nehra2010cross, shikh2007joint, kobravi2007cross, jia2020channel}. To achieve these ambitious goals, several foundational technologies have gained momentum in recent years, notably extremely large-scale antenna arrays (ELAA) \cite{wang2024tutorial}, high-frequency communications \cite{jiang2024terahertz}, i.e. millimeter wave (mmWave) and sub-terahertz (sub-THz) communications, fluid antennas (FAs)-enabled integrated communications and sensing (ISAC) \cite{yang2025fluid}, full-duplex (FD) communications \cite{smida2023full, zhang2019neural, towhidlou2017improved}, and reconfigurable intelligent surfaces (RIS) \cite{shi2024ris, fang2021secure}. However, these advances also lead to a paradigm shift in system modeling and physical-layer design, particularly due to the emergence of near-field (NF) effects arising from the expansion of antenna apertures and higher operating frequencies.

The realization of 6G scenarios typically requires the deployment of ELAAs and the utilization of high-frequency bands. As a result, the Rayleigh distance $R = \frac{2D^2}{\lambda}$, which is directly proportional to the carrier frequency $f_c$ and the squared product of the antenna aperture $D$, is correspondingly expanded by orders of magnitude. In contrast to the traditional far-field assumption, where electromagnetic wavefronts can be approximated as planar, NF communications are governed by spherical wavefronts that exhibit spatially non-stationary channel characteristics \cite{lu2024tutorial}. This fundamental change challenges traditional beamforming and spatial multiplexing designs \cite{zhang20236g}, but also opens new opportunities for focused energy transfer \cite{mou2022near}, precise user localization \cite{zhou2024near}, and improved spatial resolution \cite{ding2024utilizing}. Moreover, in \cite{sheemar2025near}, the authors investigated FD extremely large-scale multiple-input multiple-output (XL-MIMO) with reconfigurable holographic surfaces (RHS) and optimized beamforming for interference suppression. In \cite{nazar2025full}, the authors studied joint transmitting (TX)/receiving (RX) beamforming design for FD-ISAC NF systems. In \cite{du2025full}, a comprehensive overview of FD-ISAC principles, SIC techniques, receiver design, and hardware issues was provided.

Furthermore, based on the 6G imperatives mentioned above, FA-enabled systems have been proposed as a flexible solution that allows each antenna element to dynamically move within a restricted region \cite{zhu2023modeling, wong2020fluid, wu2025fas, wu2025fluid, wu2025scalable, wu2025variable, wu2025}. By exploiting location diversity, FAs can adapt to changing environments and user distributions, resulting in improved signal strength, interference avoidance, and reliability \cite{new2024tutorial}. Initial studies have demonstrated the feasibility of FAs in both single-user \cite{zhou2024near} and multi-user \cite{yang2025fluid, qin2024antenna} scenarios. Moreover, recent work has introduced rotatable antenna arrays, whose individual elements can actively adjust their three-dimensional (3D) orientations to increase the spatial degrees of freedom (DoFs) beyond what phase-only beamforming can achieve \cite{zhu2023movable}. In addition, by allowing each radiating element to move in a six-dimensional (6D) region, the 6D movable antenna (6DMA) exploits location diversity to reshape small-scale fading, strengthen desired links, and avoid interference \cite{shao20246d, shao20256d}. However, systematic methods that jointly optimize mobility for time and power control remain limited, particularly when realistic coupling and NF propagation are taken into consideration.

However, recent works on FA NF systems (FANSs) are nascent, yet rapidly consolidating around a joint-optimization paradigm that couples antenna mobility with spherical-wave NF propagation. The work in \cite{chen2024joint} investigates energy-efficiency maximization in a point-to-point NF system with multiple fixed-position antennas (FPAs) at the BS and FAs at the user side. Moreover, \cite{zhou2024near} considers a sensing-oriented FANS by integrating it into an extremely large-scale simultaneously transmitting and reflecting surface (STARS)-assisted ISAC framework and minimizing the sensing Cram\'er--Rao bounds (CRBs) via a nested block coordinate descent (BCD) algorithm. Furthermore, \cite{yang2025fluid} and \cite{yang2025enhanced} extend FANS to computing and semantic communications through the joint design of ISAC beamforming, FA positioning, and semantic extraction, showing that FA mobility improves both data rate and sensing accuracy. In parallel, \cite{zhou2024location} studies a larger-scale FANS with simultaneous transmission and reflection, where FA positioning is jointly optimized to improve location-estimation accuracy and communication performance.

Collectively, the mentioned works converge on the alternating-optimization (AO) framework to handle strongly coupled position–beamforming variables and empirically demonstrate robustness and favorable scaling. They also highlight FANS-relevant open issues, such as multi-sector operation, mobility–resource co-optimization, and system-level scaling with respect to the antennas and the users. Specifically, prior studies of FANSs typically consider half-duplex (HD) operation. However, the FD operation offers a natural companion to NF and mobility by promising near-doubling of spectral efficiency by enabling simultaneous transmission and reception over the same time and frequency resource \cite{sabharwal2014band, ju2014optimal, zhang2020ensemble, towhidlou2018adaptive}. Despite the challenges posed by strong self-interference (SI), substantial progress in analog and digital SI cancellation (SIC) has rendered the FD communication a practical candidate for access point and BS designs \cite{li2019full, chen20235g, zhang2019neural}. For the 6G network, the FD aligns with joint wireless energy transfer (WET) and communications \cite{olfat2008optimum, olfat2005optimum, shadmand2010multi}. Yet residual SI remains the primary bottleneck, especially in ELAAs, where coupling paths can be strong, and mandates co-design of array geometry, SIC, and resource allocation. Consequently, the FD becomes a powerful but delicate accelerator for the 6G when paired with NF focusing and adaptive antenna arrays.

Bridging these gaps requires coupling aperture partitioning with the FD concurrency under a realistic NF model. In dense short-range wireless-powered networks, the downlink is often used for WET while the uplink carries sensing or monitoring data, as in batteryless Internet-of-Things (IoT) and industrial deployments (e.g., factories, warehouses, and smart buildings) where frequent battery replacement is costly. Such scenarios naturally fall into the NF regime, where the channel is highly dependent on distance and geometry, thereby enabling energy focusing and throughput enhancement through spatial reconfiguration. This motivates the use of FAs, whose positions can be optimized to exploit NF propagation and improve both harvested energy and uplink throughput. Furthermore, when the BS operates in FD mode to enhance spectrum utilization, TX/RX grouping becomes important for SI management, since different group assignments lead to different physical isolation and residual leakage levels. Inspired by this, we propose a sectorization principle to FA arrays with FD transmission. Specifically, we partition the FAs aperture into multiple groups and assign each group to transmit or receive within a frame, while retaining element-level mobility within local boxes.

This multi-sector FD-FANS architecture is a natural fit for the ELAA system, as 1) grouping provides coarse spatial isolation, which is beneficial for effective SI management at the system level, 2) FD concurrency exploits shared resources for both the downlink energy transmission (DET) and the uplink data transmission (UDT), and 3) the mobility of the FAs affords extra spatial DoFs for refining NF focusing and coupling. In addition, the adopted half-TX/half-RX partition provides a balanced and practical FD sectorization that preserves sufficient spatial DoFs for both DET and UDT, while avoiding overly asymmetric designs that may degrade either TX-side energy focusing or RX-side signal collection. Compared with existing FANSs, our design closes two gaps. First, we impose geometric constraints to bound the movable regions of the FAs and introduce a half-TX/half-RX partitioning framework for a FA aperture system that is consistent with practical hardware sectorization. In addition, we jointly optimize time–power allocation, antenna positions, and FAs grouping in a FD communication system operating over a spherical-wave channel, considering both cases with and without residual SI. 

More specifically, unlike prior half-duplex FANS (HD-FANS) or fixed-aperture FD models, the binary TX/RX grouping variables in our design are not simple labels; they directly determine the TX/RX sub-apertures, alter the effective NF uplink/downlink composite gains, and affect the physical isolation that governs residual SI leakage. Meanwhile, antenna mobility influences not only desired signal enhancement but also the coupling between TX and RX sectors under spherical-wave propagation. In addition, the WET-based harvest-then-transmit protocol introduces energy-causality coupling between downlink DET and uplink UDT across time slots. As a result, time allocation, power control, antenna positioning, and TX/RX grouping are coupled through a formulation that is fundamentally different from existing FANS models with HD operation, fixed apertures, or non-sectorized transmission. The proposed joint design of antenna positioning and TX/RX grouping, together with time–power allocation under energy causality, is particularly relevant for NF wireless powered communication networks (WPCN) deployments and supports emerging applications such as dense industrial sensing, radio frequency identification (RFID)-like infrastructures, and short-range wearable/implantable monitoring.

The main contributions of the paper are outlined below. \begin{itemize}

    \item We propose an FD-FANS that jointly models spherical-wave uplink and downlink channels, residual SI after analog and digital SIC, WET, and multi-sector TX/RX grouping with element mobility and geometric constraints.
    \item We formulate a weighted sum-rate (WSR) maximization problem over time allocation, power control, antenna positioning, and binary grouping, subject to average and peak power budgets, per-antenna box limits, minimum inter-element spacing, and a half-TX/half-RX balance. The novelty here lies in the problem structure induced by the proposed FD-FANS architecture with multi-sector TX/RX partitioning, where antenna positioning, residual SI coupling, and WET-based time--power allocation are strongly intertwined. This yields a coupled formulation beyond existing FD-FANS models with fixed apertures, non-sectorized operation, or HD transmission.
    \item To tackle the resulting non-convex mixed-integer problem, we develop an efficient AO framework based on MM/SCA. The proposed algorithm is tailored to this structure and guarantees monotonic objective improvement and convergence to a stationary solution of the relaxed problem.
    \item Simulation results demonstrate that the proposed FD-FANS consistently outperforms several benchmark schemes, including HD-FANS, FD-FPA-NF system (FD-FPANS), non-grouped FD-FANS, and far-field counterparts, in terms of average sum-rate (ASR), energy efficiency (EE), and user fairness, while exhibiting robustness to residual SI and channel uncertainty. Moreover, the proposed framework exhibits near-linear scalability with respect to the number of users, the average transmit power (ATP), and the peak-to-average power ratio (PAPR), validating the effectiveness and practicality of the joint optimization design.
\end{itemize}


The remainder of this paper is organized as follows. Section~\ref{s2} presents the system models. Section~\ref{s3} formulates the joint optimization problem. Section~\ref{s4} presents the proposed solutions and algorithms, along with the complexity analysis. Simulation results and convergence analysis are shown in Section~\ref{s5}. Finally, Section~\ref{s6} concludes the paper.

\subsection*{List of Notations:} 
Capital boldface letters are used to represent matrices, and lowercase boldface letters are used to represent vectors. Scalars are represented by lower and upper-case normal fonts. $\mathbf{I}$ denotes an identity matrix. $[\;]^H$ and $[\;]^\top$ denote the Hermitian transpose and the transpose of a matrix, respectively. The $\ell_2$ norm is denoted by $\|\;\|$. $\mathcal{CN}(0,\sigma^2)$ denotes a standard complex Gaussian distribution with zero mean and variance $\sigma^2$. $\Re\!\left\{\right\}$ denotes the real-part operator. $\otimes$ denotes the Kronecker product.

\section{System Model} \label{s2}

\begin{figure}[t!]
    \centering
    \includegraphics[width=90mm]{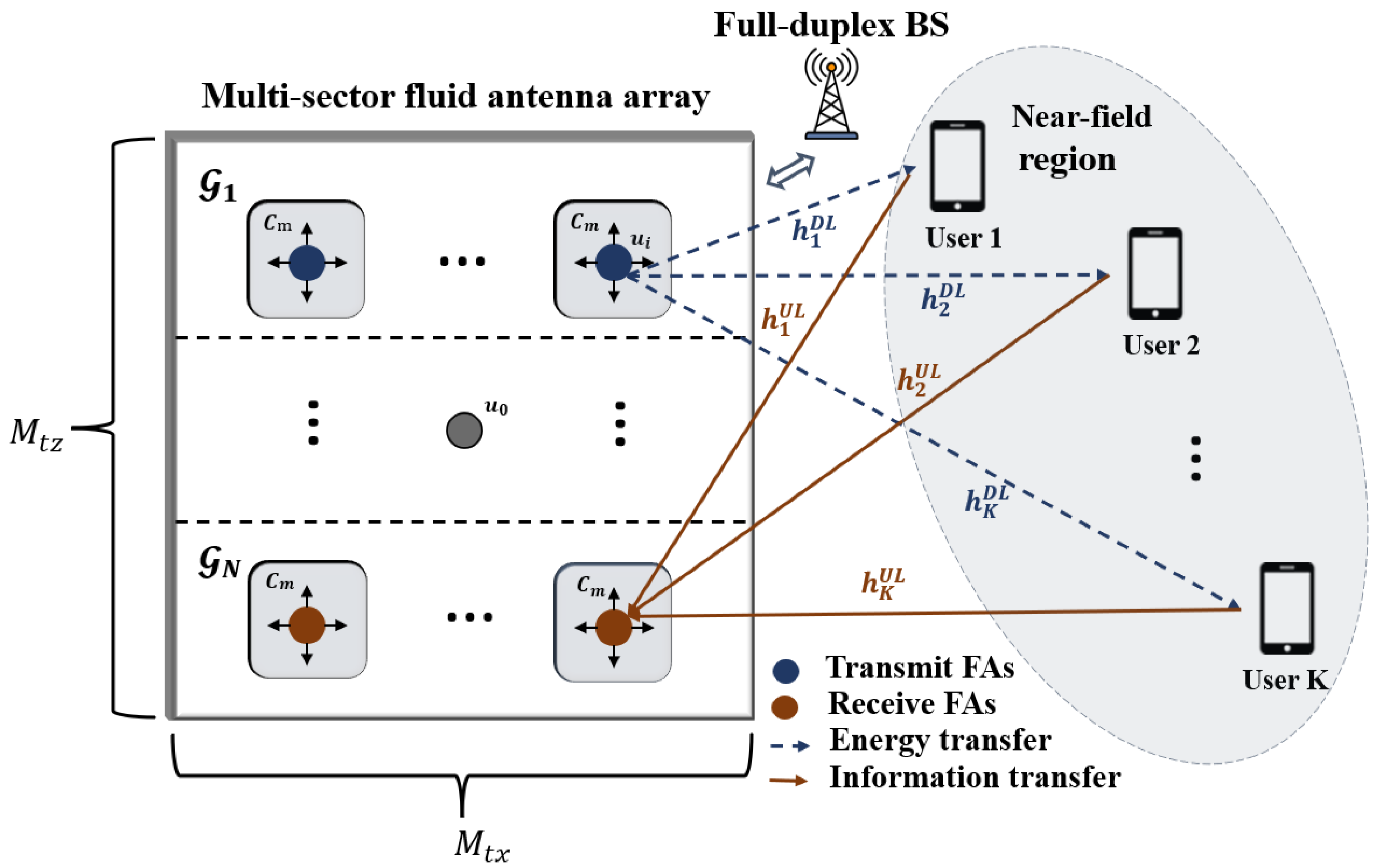}
    \caption{FD-FANS}
    \vspace{-1em}
    \label{scenario}
\end{figure}

As shown in Fig. \ref{scenario}, this paper considers a wireless communication network that consists of a BS and $K$ users, denoted by $\mathcal{K} \triangleq \{1,\ldots,K\}$, operating over the mmWave frequency band. The users are assumed to be sufficiently separated from each other and are located in the NF region of the BS. The BS is equipped with $M$ FAs, denoted by $\mathcal{M} \triangleq \{1,\ldots,M\}$ which we partition into $N$ groups, denoted by $\mathcal{N} \triangleq \{1,\ldots,N\}$, and each user terminal is equipped with a single FPA. The BS is assumed to be equipped with a stable energy supply, whereas user terminals are not endowed with dedicated energy sources. Accordingly, each user harvests energy from downlink signals of the BS, subsequently using it to sustain circuit operations and enable uplink information transmission. In addition, we consider an FD network in which the BS operates in FD mode, simultaneously transmitting downlink energy and receiving uplink data, while all users operate in time-division HD mode. Specifically, each user harvests energy during the downlink phase and transmits information in the uplink over orthogonal time slots.

\subsection{Full-Duplex Model}

\begin{figure}[t!]
    \centering
    \includegraphics[width=90mm]{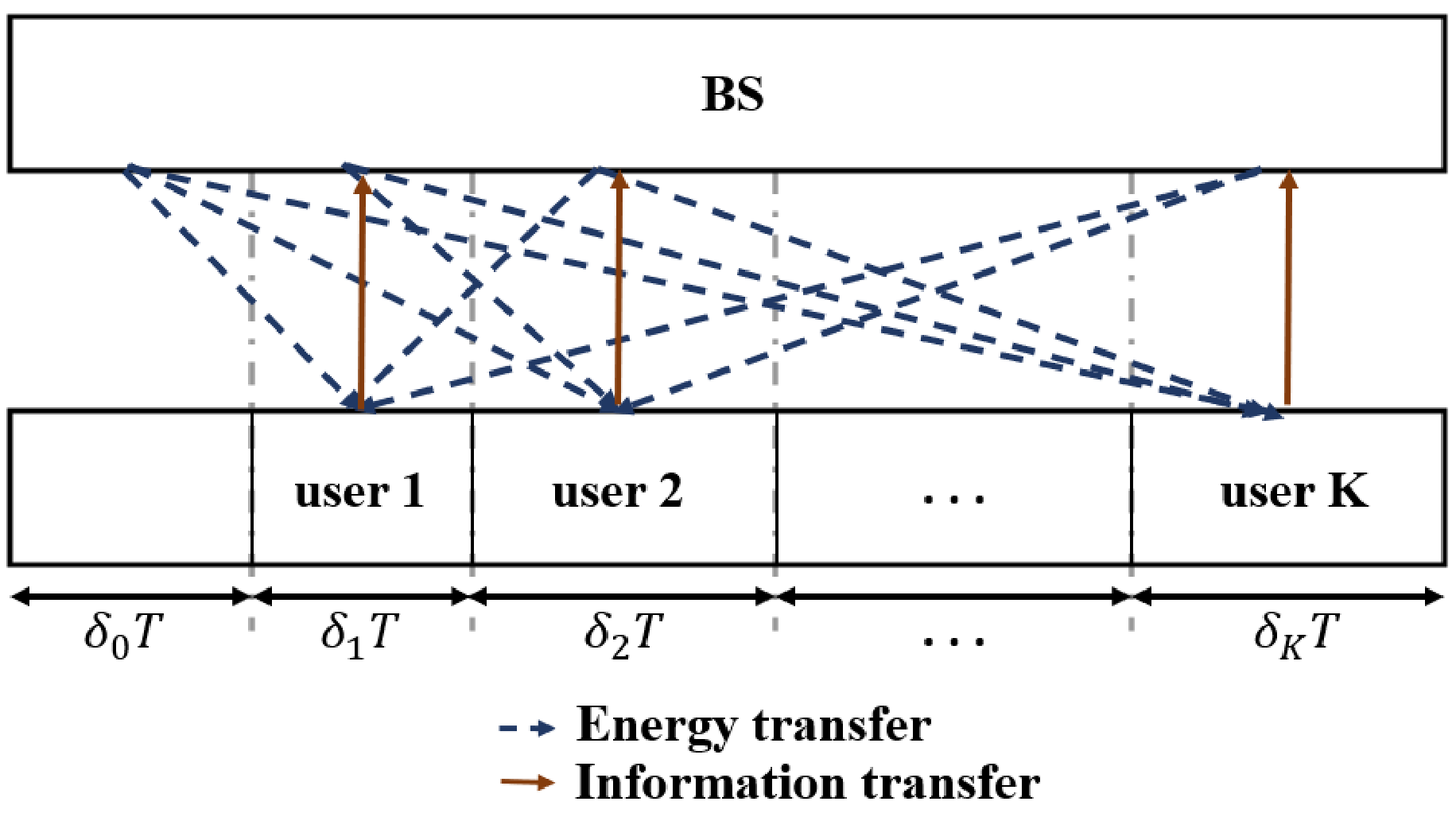}
    \caption{Energy and information transmission between the BS and users}
    \vspace{-2em}
    \label{power}
\end{figure}
We consider each communication block to span a duration of $T$, which is divided into $K+1$ distinct time slots. As shown in Fig. \ref{power}, the first slot is solely dedicated to DET. The subsequent $k$-th slot, $k \in \mathcal{K}$, is used for simultaneous DET from the BS and UDT from user $k$. We denote the time fraction allocated to slot $k$ by $\delta_k$ and the transmit power of the BS during that slot by $P_k$. To ensure practical feasibility and compliance with system-level limitations, the time-power allocation must satisfy a set of physical constraints that govern the overall transmission schedule and power budget. Thus, we let $P_{\mathrm{avg}}$ represent the ATP of the BS in all slots in each block and $P_{\mathrm{peak}}$ represent the peak transmit power of the BS in any given slot, and then we have \begin{equation} \label{eq:avg_power_constraint}
    \sum_{k=0}^{K} \delta_k P_k \leq P_{\mathrm{avg}},
\end{equation} \begin{equation} \label{eq:peak_power_constraint}
   P_k \leq P_{\mathrm{peak}}, k \in \mathcal{K}.
\end{equation}


In the uplink, users employ time division multiple access (TDMA) to send independent signals to the BS. The $k$-th slot is assigned to user $k$ for wireless information transmission. Since the BS operates in FD mode, during slot $k$ of each block, it simultaneously broadcasts energy with the TX antenna groups and receives the signal of user $k$ with the RX antenna groups. Each user $k\in \mathcal{K} \triangleq \{1,\ldots,K\}$ operates in the HD mode with a single antenna. Thus, it only transmits in its own slot and cannot harvest energy in that slot. Instead, it harvests energy from the BS during the other $K$ slots and stores it for subsequent use.

\paragraph*{Remark} TDMA is particularly suitable for the considered model because its slot-based orthogonal structure naturally matches the harvest-then-transmit protocol of energy-limited users. Since each user operates in HD mode with a single RF chain, it cannot harvest energy and transmit information simultaneously, and TDMA directly accommodates this constraint by allowing each user to harvest energy during other users’ slots and transmit only in its own slot. In addition, TDMA preserves a clear coupling among time-power allocation, harvested energy, and uplink rate, which is central to our joint optimization design. Compared with more advanced schemes such as OFDMA \cite{yao2024wireless}, it avoids introducing extra frequency-domain scheduling variables and thus leads to a more tractable formulation while remaining practically relevant for low-complexity terminals.

Furthermore, denote the transmitted signal of the BS as $x_{L,i}$, which is modeled as a BS-generated pseudo-random energy waveform used for WET with $\mathbb{E}\!\left[|x_{L,i}|^{2}\right]=1$. Let $h_{k,\mathrm{eff}}^{\mathrm{DL}}$ and $h_{k,\mathrm{eff}}^{\mathrm{UL}}$ denote the downlink and uplink complex channel coefficients between the BS and user $k$, respectively. Consider the $i$-th slot of a given block, $\forall i\in\{0,\ldots,K\}$, in which the signal received at user $k\neq i$ is \begin{equation}
  y_{k,i} = \sqrt{P_i}\, h^{\mathrm{DL}}_{k,\mathrm{eff}}\, x_{L,i} + n_{k,i},
\end{equation} where $n_{k,i}$ denote the noise of the user $k$, respectively. We further assume that $P_i$ is sufficiently large, and thus the energy harvested due to receiver noise is negligible. 

During each frame, user $k$ is allowed to transmit only in its designated slot $k$; during the remaining $K$ slots, it passively harvests energy. The energy-harvesting efficiency is modeled by $\eta_k \in (0,1)$. Then, the energy harvested by user $k \in \mathcal{K}$ from slot $k \neq i$ is \begin{equation}\label{e5}
    E_{k,i} = \eta_k P_i \left| h^{\mathrm{DL}}_{k,\mathrm{eff}}\right|^2 \delta_i T, i \in \mathcal{K},
\end{equation} and also, following the standard FD wireless-powered communication network (FD-WPCN) protocol, each user operates in time-division HD mode with a single antenna. Hence, user $k$ cannot receive/harvest energy during its own uplink transmission slot, and it harvests energy only in the remaining slots. In particular, the considered protocol does not rely on power splitting or conventional time switching; instead, users are assumed to harvest energy from the downlink waveform via separate energy harvesting (EH) circuitry, while any control signaling is negligible or separately delivered. Therefore, the total harvested energy within one communication block is \begin{equation}\label{e6}
E_{u,k} = \eta_k\left|h^{\mathrm{DL}}_{k,\mathrm{eff}}\right|^2\sum^{K}_{i=0,i\neq k}\delta_iP_i.  
\end{equation}

In addition, the energy-bearing signal $x_{L,i}$ is treated as a known waveform for WET, and users are not required to decode downlink information from $x_{L,i}$ in this work. If low-rate control signaling is needed in practice, it can be embedded in the downlink waveform or delivered via a short separate control phase/channel with negligible overhead; therefore, such overhead is omitted in~\eqref{e6}. Interested readers may refer to the literature on simultaneous wireless information and power transfer (SWIPT) \cite{an2024exploiting, zhang2013mimo, nasir2013relaying, zhang2023energy}.

Assuming a stable energy management policy, user $k$ allocates a fixed fraction $\gamma_k \in (0,1)$ of its harvested energy for data transmission in its own time slot. The corresponding ATP for user $k\in \mathcal{K}$ becomes \begin{equation}
   P_{u,k} = \frac{\gamma_kE_{u,k}}{\delta_k},
\end{equation} and then, let $x_{u,k}$ denote the normalized uplink information signal from user $k$ with $\mathbb{E}[x_{u,k}] = 0$ and $\mathbb{E}[|x_{u,k}|^2] = 1$. The received signal at the BS in time slot $k$ can be modeled as \begin{equation} \label{e8}
y_{L,k} = \sqrt{P_{u,k}}{h}_{k,\mathrm{eff}}^{\mathrm{UL}}x_{u,k} + \sqrt{P_k}h_Lx_{L,k} + n_{L,k}, k\in \mathcal{K}, \end{equation} where $x_{L,k}$ and $h_L$ represent the known pseudo-random energy signal transmitted by the BS and the loopback channel (post analog-domain SIC), respectively. Further, $n_{L,k} \sim \mathcal{CN}(0, \sigma^2)$ is the additive white Gaussian noise (AWGN) of the BS.

Although analog-domain SIC is employed, residual SI still poses a considerable challenge due to the high power of the energy signal. The effective loopback channel is expressed as \begin{equation}
    h_L = \sqrt{\phi} \hat{h}_L,
\end{equation} where $\phi$ and $\hat{h}_L$ denote the residual power level and the normalized complex coefficient of the loopback channel after analog-domain SIC, with $\mathbb{E}\!\left[|\hat{h}_L|^{2}\right]=1$. Even after analog SIC, the residual SI power can still be much stronger than the desired signal in practice, i.e., $|h^{\mathrm{UL}}_{k,\mathrm{eff}}|^{2} P_{u,k} \ll \phi P_k$. This strong SI distorts the signal received in \eqref{e8} due to the limited TX/RX dynamic ranges, creating additional noise whose power scales with the TX/RX power of the BS. For simplicity, we assume the ideal automatic gain control (AGC) and an infinite transmitter dynamic range as in \cite{ju2014optimal}. Therefore, the limited dynamic range of the receiver introduces quantization noise after analog-to-digital conversion (ADC), which is modeled as $n_{q,k} \sim \mathcal{CN}(0, \zeta \sigma^2_q)$ where $\zeta \ll 1$ and $\sigma_q^2 = \mathbb{E}\left[ \left|y_{L,k} \right|^2\right ] = P_{u,k}\left |h^{\mathrm{UL}}_{k,\mathrm{eff}}\right|^2 + \phi P_k + \sigma^2, k\in \mathcal{K}$. 

Furthermore, although the estimation errors of channels $h^{\mathrm{UL}}_{k,\mathrm{eff}}$ and $h^{\mathrm{DL}}_{k,\mathrm{eff}}$ are negligible, the error in estimating $h_L$ cannot be ignored because the SI at the BS is much stronger than the desired signal. Thus, by letting $ \bar{h}_L$ represent the estimated loopback channel, we have \begin{equation}
    \hat{h}_L = \bar{h}_L + \sqrt{\epsilon} \tilde{h}_L,
\end{equation} where $ \sqrt{\epsilon} \tilde{h}_L$ denotes the channel estimation error with $\epsilon \ll 1$ and $\tilde{h}_L \sim \mathcal{CN}(0,1)$. Therefore, the received signal after ADC and digital domain SIC can be given by\begin{equation}\label{e11}
\begin{split}   \bar{y}_{L,k} & = y_{L,k} + n_{q,k} - \sqrt{\phi P_k} \bar{h}_L x_{L,k} \\ & = \sqrt{P_{u,k}}h^{\mathrm{UL}}_{k,\mathrm{eff}}x_{u,k} + \sqrt{\epsilon \phi P_k} \tilde{h}_L x_{L,k} + n_{q,k} + n_{L,k}, k\in \mathcal{K}.\end{split}\end{equation}

Consequently, the signal-to-interference-plus-noise ratio (SINR) at the BS of user $k$ is derived as \begin{equation} \label{e12}\varrho_k = \frac{\beta_kH_{k,\mathrm{eff}}}{(\tau P_k + \sigma^2)\delta_k}\sum^{K}_{i=0,i\neq k}\delta_iP_i, k\in \mathcal{K},
\end{equation} where $\beta_k = \gamma_k \eta_k$ denotes the combined energy conversion and transmission efficiency, $H_{k,\mathrm{eff} = |h^{\mathrm{UL}}_{k,\mathrm{eff}}|^2 |h^{\mathrm{DL}}_{k,\mathrm{eff}}|^2}$ is the uplink and downlink composite channel gain, and $\tau P_k$ denotes the power of effective SI at the $i$-th slot with \begin{equation} \label{tau}
    \tau = \phi(\epsilon + \zeta),
\end{equation} where $\phi$ denotes the residual loopback ratio after analog SIC. The term $\epsilon$ accounts for the dominant digital-domain residual caused by imperfect loopback channel estimation, while $\zeta$ captures the receiver-side impairment due to quantization noise/finite dynamic range when strong SI drives the ADC close to saturation. Hence, the factor $\tau$ captures the effective residual SI power after the cascade of analog and digital SIC, and appears in the SINR expression \eqref{e12} as the term $\tau P_k$, reflecting the fact that the residual SI-related impairment scales linearly with the BS transmit power $P_k$.

It is worth noting that the factor $\tau$ is physically related to the residual loopback ratio $\phi$. In practical FD arrays operating in the NF region, $\phi$ depends on the electromagnetic coupling and isolation between the TX and RX sub-arrays, which are determined by the antenna locations and the TX/RX partitioning. Intuitively, better physical isolation, such as larger TX/RX separation or weaker mutual coupling, reduces loopback leakage and results in a smaller $\phi$ (and hence a smaller $\tau$), whereas closer spacing or unfavorable placement leads to stronger leakage and a larger $\phi$. Moreover, NF beam focusing reshapes the spatial field and energy distribution around the array, which may further affect the amount of leakage captured by the RX sub-array and thus strengthen the dependence of $\phi$ on the antenna locations.

The adopted SI representation is an effective system-level abstraction. In practice, the residual leakage may depend more explicitly on the TX/RX geometry, antenna positions, and hardware impairments. In this paper, these effects are absorbed into the effective parameter $\tau$ for analytical tractability, and the impact of antenna positioning and grouping on SI mitigation is interpreted through this effective parameterization.

Finally, the achievable data rate for user $k\in \mathcal{K}$ under the FD model is \begin{equation}\label{e14}
\begin{split}     
R_k (\boldsymbol{\delta},\boldsymbol{P}) & = \delta_k\log_2(1 + \frac{\varrho_k}{\Upsilon})\\ & = \delta_k\log_2\left(1 +  \frac{\beta_k H_{k,\mathrm{eff}}}{\Upsilon(\tau P_k + \sigma^2)}\frac{1}{\delta_k}\sum^{K}_{i=0,i\neq k}\delta_iP_i\right),
\end{split}
\end{equation} where $\boldsymbol{\delta} = [\delta_0, \delta_1, ...\delta_K]$, $\mathbf{P} = [P_0, P_1, ...P_K]$, $\Upsilon > 1$ captures the SINR gap due to practical modulation schemes.

\vspace{-0.8em}
\subsection{Multi-Sector FANS System}

\begin{figure}[t!]
    \centering
    \includegraphics[width=90mm]{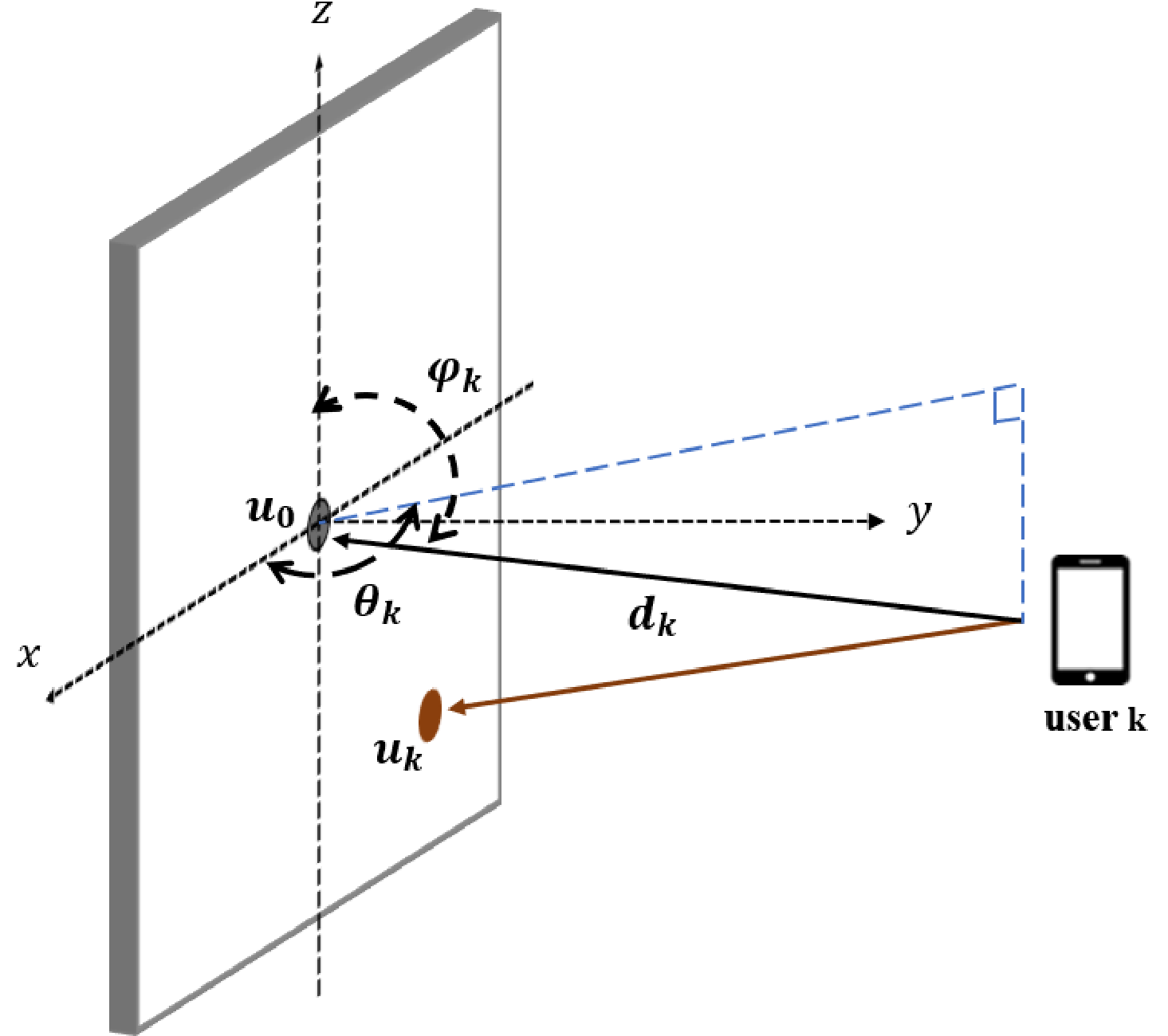}
    \caption{Geometric coordinates of user $k$ and the FAs}
    \vspace{-2em}
    \label{fa}
\end{figure}

To further enhance spatial resolution and accommodate NF propagation characteristics, we consider a BS architecture equipped with two-dimensional (2D) reconfigurable FA arrays. The TX/RX arrays are composed of $M_{tx} \times M_{tz}$ and $M_{rx} \times M_{rz}$ elements, respectively. These elements are arranged in a uniform planar array (UPA) and can dynamically reposition within predetermined rectangular movement zones. As shown in Fig. \ref{fa}, the spatial location of the $m$-th FA element is described by $\mathbf{u}_m=[x_m, z_m]$, $m\in\mathcal{M}$. The element is allowed to move only within a bounded rectangular region
$\mathcal{C}_m=[x_m^{\min},x_m^{\max}]^\top\times[z_m^{\min},z_m^{\max}]$,
where $(x_m^{\min},x_m^{\max})$ and $(z_m^{\min},z_m^{\max})$ denote the fixture limits along the $x$- and $z$-axes, respectively. This yields the region constraint $\mathbf{u}_m\in\mathcal{C}_m$. Moreover, to prevent antenna overlaps and avoid excessively strong NF mutual coupling when the elements move within the aperture, we impose a minimum inter-element spacing constraint $\| \mathbf{u}_m - \mathbf{u}_i \|\ge d_{\mathrm{min}}$ for all $m\neq i$, where $d_{\mathrm{min}}$ is typically chosen according to the element size. The centroid of the transmit antenna array is fixed and serves as the reference point for geometric computations. 

The BS simultaneously serves $K$ users, each equipped with a single FPA. To maintain adequate spatial DoFs for beamforming and interference management, we assume $K \leq M_{tx} \times M_{tz}$. The location of user $k \in \mathcal{K}$ is represented by \begin{equation}
     \mathbf{b}_k = [d_k\cos(\theta_k) \sin (\varphi_k), d_k\sin(\theta_k) \sin (\varphi_k), d_k\cos(\varphi_k)],
\end{equation} where $\theta_k$ and $\varphi_k$ denote the azimuth and elevation (broadside) angles of user $k$, respectively, and $d_k$ denotes the distance of user $k$ from the antenna array center of the BS. 

Further, the received signal $y_k$ in the BS can be expressed as \begin{equation}
  y_k = \rho_k\mathbf{a}^H_k(\theta_k, \varphi_k, d_k, \mathbf{u})\mathbf{X} + \mathbf{n}_{k} = \mathbf{h}^{\mathrm{UL}}_k(\mathbf{u})^H \mathbf{X} + \mathbf{n}_{k}, k\in\mathcal{K},
\end{equation} where $\rho_k$ captures the large-scale path loss, $\mathbf{a}_k(\cdot)$ denotes the composite steering vector that accounts for NF spherical wavefronts, $\mathbf{h}^{\mathrm{UL}}_k(\mathbf{u})$ denotes the NF uplink channel vector, $\mathbf{X}$ is the transmitted signal vector, and $n_k \sim \mathcal{CN}(0, \sigma^2_u
 \mathbf{I})$ is the AWGN.

Due to the proximity of the users, the spherical nature of wave propagation must be taken into account. The total steering vector can be given by \begin{equation}
    \mathbf{a}_t(\theta_k, \varphi_k, d_k, \mathbf{u}) = \mathbf{a}_{tx}(\theta_k, \varphi_k, d_k, \mathbf{x}_t) \otimes \mathbf{a}_{tz}(\varphi_k, d_k, \mathbf{z}_t), k\in \mathcal{K},
\end{equation} where $\mathbf{x}_t$ and $\mathbf{z}_t$ represent the arrays of FA coordinates along the $x$ and $z$ axes. Each element in the steering vectors is calculated under a non-uniform spherical wave (NUSW) model \cite{liu2023near}. Specifically, for user $k \in \mathcal{K}$, we have \begin{align}\label{e17}
    & \mathbf{a}_{tx}(\theta_k, \varphi_k, d_k, x_m) = \notag \\ & \quad e^{j\frac{2\pi}{\lambda}\left(x_m\cos(\theta_k)\sin(\theta_k)-\frac{x_m^2(1-\cos^2(\theta_k)\sin^2(\varphi_k))}{2d_k}\right)}, x_m \in \mathbf{x}_t, 
\end{align} and \begin{align} \label{e18}
       & \mathbf{a}_{tz}(\varphi_k, d_k, z_m)  = \notag \\ & \quad e^{j\frac{2\pi}{\lambda}\left(z_m\cos(\theta_k)-\frac{z_m^2\sin^2(\varphi_k))}{2d_k}\right)}, z_m \in \mathbf{z}_t, m\in \mathcal{M},\end{align} which are derived from a second-order Taylor expansion of the exact spherical-wave propagation distance, i.e., they correspond to a Fresnel/parabolic approximation rather than a full-wave electromagnetic formulation. Accordingly, the model accuracy is determined by the first neglected higher-order phase term in the distance expansion. As commonly adopted in the NF studies (see \cite{liu2023near} and \cite{yang2025fluid}), the second-order approximation is regarded as valid when the omitted higher-order phase contribution satisfies the standard Fresnel truncation criterion, e.g., the classical $\frac{\pi}{8}$ phase-error bound. We thus use \eqref{e17} and \eqref{e18} as a tractable system-level NF channel model for the subsequent joint optimization.

\subsection{Multi-Group Antenna Array}

We extend the proposed FD-FANS by partitioning the UPA of the BS, consisting of $M$ antennas, into $N$ disjoint groups $\mathcal{G}_n \subseteq \mathcal{M}, \forall n \in \mathcal{N}$. We assume that each antenna is assigned to exactly one group, i.e., $\bigcup_{n=1}^{N}\mathcal{G}_n = \mathcal{M}$, and the groups are mutually disjoint, i.e., $\mathcal{G}_n\cap\mathcal{G}_{n'} = \emptyset, \forall n\neq n'$. Additionally, each group is assigned exclusively to either TX or RX within each communication interval. The selection set of group $n \in \mathcal{N}$ is thus given by 
\begin{equation}
    \mathcal{G}_n = \{g\times(n-1)+1, \dots, g\times n\},
\end{equation} where $g = M/N$ denotes the group size. 

Then, we define the binary group selection vector as $\mathbf{z} = [z_1, z_2, \dots, z_N]^\top \in \{0,1\}^N$, where $z_n = 1$ indicates that group $\mathcal{G}_n$ is selected as a TX group, and $z_n = 0$ indicates a RX group. The binary variable $z_n$ is introduced to characterize a practical multi-sector FD-FANS architecture, where the grouping scheme acts as a system-level mechanism for SI management in FD mode. Different TX/RX group assignments induce different relative geometries and physical isolation levels between the TX and RX sub-arrays, which in turn result in different effective residual SI leakage levels. Therefore, the grouping decision directly impacts the residual SI coupling, and consequently, the achievable rate and the subsequent joint resource-allocation design.

Let $\mathbf{S}_{\mathrm{TX}}(\mathbf{z})\in \left\{ 0,1 \right\}^{M_{tx}\times M}$ and $\mathbf{S}_{\mathrm{RX}}(\mathbf{z})\in \left\{ 0,1 \right\}^{M_{rx}\times M}$ denote the selection matrices that index the TX and RX antennas, respectively. The composite channel gain is thus defined as
\begin{equation}
\tilde{\mathbf{H}}_k(\mathbf{u},\mathbf{z}) \triangleq \mathbf{G}_{U,k}(\mathbf{u},\mathbf{z})\,\mathbf{G}_{D,k}(\mathbf{u},\mathbf{z}),
\label{e20}
\end{equation} with \begin{align}
\mathbf{G}_{U,k}(\mathbf{u},\mathbf{z}) 
&= \big\|\,\mathbf{S}_{RX}(\mathbf{z})\,\mathbf{h}^{\mathrm{UL}}_k(\mathbf{u})\big\|_2^{2}\notag\\
& = \sum_{n=1}^{N} (1 - z_n)\sum_{m \in \mathcal{G}_n} \big|h_{k,m}^{\mathrm{UL}}(\mathbf{u})\big|^2, k\in \mathcal{K}, \label{e21}\\
\mathbf{G}_{D,k}(\mathbf{u},\mathbf{z}) 
&= \big\|\,\mathbf{h}^{\mathrm{DL}}_k(\mathbf{u})\,\mathbf{S}_{TX}(\mathbf{z})\big\|_2^{2}\notag\\
&=\sum_{n=1}^{N} z_n
\sum_{m \in \mathcal{G}_n} \big|h_{k,m}^{\mathrm{DL}}(\mathbf{u})\big|^2, k\in \mathcal{K},\label{e22}
\end{align}where $\mathbf{h}^{\mathrm{UL}}_{k,m}(\mathbf{u}) = \big[h_{k,1}^{\mathrm{UL}}(\mathbf{u}),\dots,h_{k,M}^{\mathrm{UL}}(\mathbf{u})\big]^\top$ and $\mathbf{h}^{\mathrm{DL}}_{k,m}(\mathbf{u}) = \big[h_{k,1}^{\mathrm{DL}}(\mathbf{u}),\dots,h_{k,M}^{\mathrm{DL}}(\mathbf{u})\big]^\top$ are the NF uplink and downlink channel vectors. 

To additionally assess the impact of imperfect channel state information (CSI) in a simple yet tractable manner, we introduce an effective CSI uncertainty factor $\xi \in [0,1)$ and define the degraded composite gain as
\begin{equation}
\mathbf{H}_k(\mathbf{u},\mathbf{z}) = (1-\xi)\,\tilde{\mathbf{H}}_k(\mathbf{u},\mathbf{z}), k \in \mathcal{K},
\end{equation} where $\xi$ captures the effective loss in the composite NF gain due to imperfect uplink/downlink CSI acquisition, channel estimation mismatch, and related modeling errors. This simplified uncertainty modeling is also motivated by prior robust beamforming studies under imperfect CSI in RIS-aided systems \cite{yao2023robust}. Therefore, the achievable rate of user $k\in\mathcal{K}$ is expressed as \begin{equation}
R_k(\boldsymbol{\delta},\mathbf{P},\mathbf{u},\mathbf{z})
= \delta_k \log_2\!\left(
1+\frac{\beta_k\,\mathbf{H}_k(\mathbf{u},\mathbf{z})}{\Upsilon\big(\tau P_k+\sigma^2\big)\delta_k}\sum_{i=0,\;i\neq k}^{K}\delta_i P_i
\right).
\label{eq:rate}
\end{equation}

\paragraph*{Remark}In practice, the residual leakage depends on the TX/RX geometry and partition, and thus could be more explicitly modeled through a position-dependent loopback channel $H_L(\mathbf{u},\mathbf{z})$. Such a model may further include frequency selectivity and hardware nonlinearities. To maintain analytical tractability and focus on communication-level joint optimization, this paper does not explicitly derive $H_L(\mathbf{u},\mathbf{z})$, but instead captures its net effect through the effective parameter $\tau$.

\section{Problem Formulation}\label{s3}

In this section, an optimization problem is formulated to maximize the uplink WSR of the proposed FD-FANS. To fully unlock the potential of antenna mobility, we aim to jointly optimize the time allocation vector $\boldsymbol{\delta} =[\delta_0,\ldots,\delta_K]$, and the power allocation vector $\mathbf{P}=[P_1,\ldots,P_K]$, the FAs position vector $\mathbf{u}=[u_1,\ldots,u_M]$, and the group selection vector $\mathbf{z}=[z_1,\ldots,z_N]$. The problem is formulated as \begin{subequations} \label{e23}
\begin{align}
\max_{\boldsymbol{\delta}, \mathbf{P}, \mathbf{u}, \mathbf{z}}\; &\sum^K_{k=1}\alpha_kR_k(\boldsymbol{\delta}, \mathbf{P}, \mathbf{u}, \mathbf{z}) \ \label{eq:objective}\\
\textrm{s.t.} \quad & \eqref{eq:avg_power_constraint}, \eqref{eq:peak_power_constraint}, \notag \\
& \sum_{k=0}^{K} \delta_k \leq 1,\label{eq:time_constraint}\\
& \sum_{n=1}^N z_n = \frac{N}{2}, \label{eq:group_constraint}\\
& \mathbf{u}_m \in \mathcal{C}_m, m \in \mathcal{M}, \label{eq:position_constraint} \\
& \| \mathbf{u}_m - \mathbf{u}_i \|_2 \ge d_{\mathrm{min}}, i \in \mathcal{M}, i \neq m, \label{eq:separation_constraint} \\
& 0 \le \delta_{k} \le 1, P_{k} \ge 0, k \in \mathcal{K}, \label{eq:var_range_continuous} \\
& z_n \in \{0, 1\}, n \in \mathcal{N}, \label{eq:var_range_discrete}
\end{align}
\end{subequations} where $\alpha_{k}$ is the non-negative rate weight for user $k$. \eqref{eq:avg_power_constraint} ensures that the ATP is limited by $P_{\mathrm{avg}}$, while \eqref{eq:peak_power_constraint} restricts the instantaneous TX power by $P_{\mathrm{peak}}$. \eqref{eq:time_constraint} guarantees that the total time allocation within one block does not exceed unity. \eqref{eq:group_constraint} enforces the FD operation by requiring half of the antenna groups to function as transmitters and the other half as receivers. \eqref{eq:position_constraint} restricts the movement of each FA within its predefined region $\mathcal{C}_m$, while \eqref{eq:separation_constraint} imposes a minimum separation distance $d_{\mathrm{min}}$ to avoid collision and NF coupling. Finally, \eqref{eq:var_range_continuous} and \eqref{eq:var_range_discrete} specify the feasible domains of the optimization variables. The WSR maximization problem is a strongly coupled non-convex mixed-integer problem due to the spherical-wave channel, residual SI, antenna mobility constraints, and binary TX/RX grouping. Since global optimization is computationally intractable, we instead develop efficient AO/MM-SCA methods that ensure monotonic improvement and convergence to a stationary point of the relaxed problem.

\section{Proposed Joint Optimization Algorithm}\label{s4}

To tackle the resulting non-convex problem, we develop an efficient AO framework by decomposing the variables into three blocks, namely $\{\boldsymbol{\delta}, \mathbf{P}\}$, $\{\mathbf{u}\}$, and $\{\mathbf{z}\}$, corresponding to time–power allocation, antenna positioning, and group selection, respectively. These blocks are iteratively optimized while keeping the others fixed. The resulting subproblems and their solutions are detailed in the following, along with the overall algorithm and its complexity analysis.

\subsection{Time and Power Allocation}

For any given $\{\mathbf{u}\}$ and $\{\mathbf{z}\}$, the subproblem with respect to $\{\boldsymbol{\delta}, \mathbf{P}\}$ is given by \begin{align} \label{e25}
&\max_{\boldsymbol{\delta}, \mathbf{P}}\
\sum^K_{k=1}\alpha_kR_k(\boldsymbol{\delta}, \mathbf{P}) \\
&\textrm{s.t.} \
\eqref{eq:avg_power_constraint}, \eqref{eq:peak_power_constraint}, \eqref{eq:time_constraint}, \eqref{eq:var_range_continuous}. \notag
\end{align}

In the following, we first study the time–power allocation for WSR maximization in \eqref{e23} under perfect SIC, i.e., $\tau = 0$ in \eqref{e14}, and then extend the analysis to the general case with finite SI, i.e., $\tau > 0$. 

\subsubsection{Perfect SIC ($\tau = 0$)}

We begin by examining the case where the SI at the BS is perfectly suppressed. In this ideal regime, the uplink signals from the users are not affected by loop interference. The resulting SINR for the user $k \in \mathcal{K}$ can be written as \begin{equation}
\varrho_k = \frac{\beta_k H_k(\mathbf{u},\mathbf{z})}{\sigma^2\delta_k}
\sum_{\substack{i=0 \\ i\neq k}}^{K}\delta_i P_i,
\end{equation} and thus, the corresponding achievable rate is \begin{equation}
R_k^{(\tau = 0)}(\boldsymbol{\delta},\mathbf{P})
=\delta_k \log_2\!\left(
1+\frac{\beta_k H_k(\mathbf{u},\mathbf{z})}{\Upsilon\sigma^2\delta_k}
\sum_{\substack{i=0 \\ i\neq k}}^{K}\delta_i P_i
\right).
\end{equation}

To simplify the non-convex average power constraint, we introduce energy variables $E_k=\delta_k P_k, k \in \mathcal{K}$, as in \cite{orhan2014energy}, where $E_k$ denotes the energy used by the BS in slot $k$, leading to \begin{equation}\label{e28}
\sum_{k=0}^{K}E_k \leq P_{\mathrm{avg}},
\end{equation}
\begin{equation} \label{e29}
E_k - P_{\mathrm{peak}}\delta_k \leq 0, k \in \mathcal{K}.
\end{equation} 

Accordingly, the achievable rate of user $k\in \mathcal{K}$ becomes \begin{equation}
R_k(\boldsymbol{\delta},\mathbf{E})
=\delta_k \log_2\!\left(
1+\frac{\omega_k}{\delta_k}
\sum_{\substack{i=0 \\ i\neq k}}^{K}E_i
\right),
\end{equation} where $\mathbf{E}=[E_0, E_1, \dots, E_K]$, $\omega_k = \frac{\beta_k H_k(\mathbf{u},\mathbf{z})}{\Upsilon\sigma^2}$. Thus, the optimization problem can be expressed equivalently as 
\begin{subequations} \label{e30}
\begin{align}
\max_{\boldsymbol{\delta}, \mathbf{E}}\; &\sum^K_{k=1}\alpha_kR_k(\boldsymbol{\delta}, \mathbf{E}) \ \\
\textrm{s.t.} \quad &  \eqref{eq:time_constraint}, \eqref{e28}, \eqref{e29}, \notag \\
& 0 \le \delta_{k} \le 1, E_{k} \ge 0, k \in \mathcal{K},\label{e31.1}
\end{align}
\end{subequations} in which the original problem \eqref{e25} is converted to a joint time-power allocation problem. Further, we note that the objective function of \eqref{e30} increases monotonically with $E_k, k\in \mathcal{K}$, for any given $\boldsymbol{\delta}$, and consequently, constraint \eqref{e28} should hold equality. Therefore, we can obtain the optimal solution for \eqref{e30} by solving the problem \begin{subequations} \label{e31}
\begin{align}
\max_{\boldsymbol{\delta}, \mathbf{E}}\; &\sum^K_{k=1}\alpha_kR_k(\boldsymbol{\delta}, \mathbf{E}) \ \\
\textrm{s.t.} \quad &  \eqref{eq:time_constraint}, \eqref{e29}, \eqref{e31.1} \notag \\ 
& \sum_{k=0}^{K}E_k = P_{\mathrm{avg}},
\end{align}
\end{subequations} with $R_k(\boldsymbol{\delta},\mathbf{E})
=\delta_k \log_2\!\left(
1+\frac{\omega_k}{\delta_k}(P_\mathrm{avg}-E_k)\right)$. As proved in \cite{ju2014optimal1}, the logarithmic function  $R_k(\boldsymbol{\delta},\mathbf{E})$ is a concave function of $\boldsymbol{\delta}$ and $\mathbf{E}$. The problem \eqref{e31} is therefore convex, and thus the Lagrangian of problem \eqref{e31} is given by \begin{align}
&\mathcal{L}(\boldsymbol{\delta},\mathbf{E},\Lambda,\nu)= \notag \\ & \quad \sum_{k=1}^K \alpha_k R_k(\boldsymbol{\delta},\mathbf{E})
-\Lambda\left(\sum_{k=0}^K \delta_k-1\right) +\nu\left(\sum_{k=0}^K E_k-P_{\mathrm{avg}}\right), \label{e32}
\end{align} where $\Lambda \geq 0$ and $\nu$ are the Lagrange multipliers. Furthermore, the associated dual function is \begin{equation}
\mathcal{Q}(\Lambda,\nu)=\max_{(\boldsymbol{\delta},\mathbf{E})\in\mathcal{D}}
\;\mathcal{L}(\boldsymbol{\delta},\mathbf{E},\Lambda,\nu),
\label{eq:Dual}
\end{equation} with $\mathcal{D}$ defined by the feasibility region. For given $\Lambda \ge 0$, $\nu$ and $\alpha_k > 0, k \in \mathrm{K}$, the maximizer of the Lagrangian \eqref{e32} can be obtained by \begin{align}
\hat{\delta_0}&=\begin{cases}
1,& \nu>0 \ \text{and}\ (-\Lambda+\nu P_{\mathrm{peak}})>0,\\
0,& \text{otherwise},
\end{cases} \label{eq:tau0}\\[1ex]
\hat{E}_0&=\begin{cases}
P_{\mathrm{peak}}\hat{\delta_0},& \nu>0,\\
0,& \text{otherwise},
\end{cases} \label{eq:E0}\\[1ex]
\hat{\delta_k}&=\min\!\left[\Big(\tfrac{\omega_k}{q_k^*}(P_{\mathrm{avg}}-\hat{E}_k)\Big)^{+},\,1\right],  k \in \mathrm{K}, \label{eq:taui}\\
\hat{E}_k&=\min\!\left[\Big(P_{\mathrm{avg}}+\tfrac{\hat{\delta_k}}{\omega_k}
-\tfrac{\alpha_k \hat{\delta_k}}{\nu \ln 2}\Big)^{+},\,P_{\mathrm{peak}}\hat{\delta_k}\right], k \in \mathrm{K}, \label{eq:Ei}
\end{align} with $(x)^{+} \triangleq \max (0, x)$, and $q_k^*$ is the solution of $f(q_k)=\frac{\Lambda^*\ln 2}{\alpha_k}$, where $f(q) \triangleq \ln(1+q)-\frac{q}{1+q}$, as proved in the Appendix. The optimal solution $\boldsymbol{\delta}^* = [\delta_0^*, \delta_1^*, ...\delta_K^*]$ and $\mathbf{E}^* = [E_0^*, E_1^*, ..., E_K^*]$ can be obtained by alternating updates of $\{\delta_k\}$ and $\{E_k\}$ using \eqref{eq:taui} and \eqref{eq:Ei}, followed by the computation of $\hat{\delta_0}$ and $\hat{E}_0$ using \eqref{eq:tau0} and \eqref{eq:E0}. Once $(\boldsymbol{\delta}^*,\mathbf{E}^*)$ is obtained, the optimal power allocation $\mathbf{P}^*$ follows as \begin{equation}
    P_k^*=\frac{E_k^*}{\delta_k^*}.
\end{equation}

Furthermore, the outer minimization of $\mathcal{Q}(\Lambda,\nu)$ is handled by sub-gradient methods, e.g. the ellipsoid algorithm, with subgradients \begin{align} \label{e39}
\mu_\Lambda &=\sum_{k=0}^K \delta_k^*-1,\\
\mu_\nu &=P_{\mathrm{avg}}-\sum_{k=0}^K E_k^*.
\label{e40}
\end{align}

The algorithm to solve problem \eqref{e25} for the perfect SIC case is presented in \textbf{Algorithm \ref{a1}}. In the perfect-SIC scenario, problem \eqref{e25} is convex and is solved by the proposed dual decomposition method. Since the updates of $\delta_k$ and $E_k$ across all users require $O(K)$ operations in total, whereas the updates of $\delta_0$, $E_0$, and $(\Lambda,\nu)$ only involve constant complexity, the overall per-iteration complexity is $O(K)$. The dual iteration is terminated when the subgradient norm satisfies $\left\|\big[\mu_\Lambda,\mu_\nu\big]\right\|_2 \le \varepsilon,$ or when the relative change of the dual objective falls below $\varepsilon$. In our implementation, the ellipsoid-based dual update generally converges within 30 iterations under the considered simulation settings.

\begin{algorithm}[t]
\caption{Time-Power Allocation ($\tau = 0$) }
\label{a1}
\begin{algorithmic}[1]
\STATE \textbf{Input:} $s=0$, $s_{max}=50$, $\varepsilon=10^{-4}$ and dual variables $(\Lambda,\nu)$.
\REPEAT
    \STATE \textbf{Update $\delta_k$ and $E_k$:} 
    \FOR{$k=1$ to $K$}
        \STATE Update $\delta_k^{(s+1)}$ and $E_k^{(s+1)}$ using closed-form expressions in \eqref{eq:taui} and \eqref{eq:Ei}.
    \ENDFOR
    \STATE \textbf{Update $\delta_0$ and $E_0$:} Compute $\delta_0^{(s+1)}$ and $E_0^{(s+1)}$ via \eqref{eq:tau0} and \eqref{eq:E0}.
    \STATE \textbf{Dual Update:} Update $(\Lambda,\nu)$ using subgradient method with subgradients in \eqref{e39} and \eqref{e40}.
    \STATE $s \gets s+1$.
\UNTIL{$\left\|\big[\mu_\Lambda,\mu_\nu\big]\right\|_2 \le \varepsilon$, the relative change of the dual objective falls below $\varepsilon$, or $s \ge s_{\max}$ }
\STATE \textbf{Output:} Optimal solution $(\boldsymbol{\delta}^*,\mathbf{E}^*)$; obtain $P_k^* = E_k^*/\delta_k^*$.
\end{algorithmic}
\end{algorithm}

\subsubsection{Imperfect SIC ($\tau>0$)}

In the presence of residual SI of the BS, problem \eqref{e25} is non-convex. We adopt an alternating scheme initialized at the perfect-SIC solution $(\boldsymbol{\delta}^*,\mathbf{P}^*)$, and iteratively update time and power until the WSR no longer improves. To simplify the time update, we use the sum-rate obtained by enforcing the average power constraint with equality as \begin{align}
&R_{k}^{(\tau > 0)}(\delta_k,P_k)
= \notag \\ & \quad \delta_k \log_{2}\!\left(
1+\frac{\beta_k H_k(\mathbf{u}, \mathbf{z})}{\Upsilon(\tau P_k+\sigma^2)}\cdot
\frac{P_{\mathrm{avg}}-\delta_k P_k}{\delta_k}
\right).
\label{eq:rfsi_sur}
\end{align}

We note that $\boldsymbol{\delta}^{s}$ for a given $\mathbf{P}^{(s-1)}$ can be obtained by solving \begin{subequations} \label{e42}
\begin{align}
\max_{\boldsymbol{\delta}, \mathbf{P}}\; &\sum_{k=1}^{K}\alpha_kR_k^{(\tau > 0)} \!\big(\delta_k,P_k^{(s-1)}\big)\ \\
\textrm{s.t.} \quad &  \eqref{eq:time_constraint}, \notag \\
& \sum_{k=0}^{K}P_k^{(s-1)}\delta_k\le P_{\mathrm{avg}},\\
& \delta_k \geq 0, k\in \mathcal{K}, \label{e42.1}
\end{align}
\end{subequations} where $R_k^{(\tau > 0)} \!\big(\delta_k,P_k^{(s-1)}\big)$ is a concave function given $\mathbf{P}=\mathbf{P}^{(s-1)}$ \cite{ju2014optimal1}, and can be solved by Lagrangian duality. The Lagrangian is given by \begin{align}
& \mathcal{L}^{(\tau>0)}(\delta,\Lambda,\nu)=  \sum_{k=1}^{K}\alpha_k\,R_{k}^{(\tau >0)}(\delta_k,P_k^{(s-1)}) - \notag \\ & \quad\Lambda\!\left(\sum_{k=0}^{K}\delta_k-1\right)
- \nu\!\left(\sum_{k=0}^{K}P_k^{(s-1)}\delta_k-P_{\mathrm{avg}}\right),
\label{eq:L_fsi}
\end{align} and therefore, the dual problem of \eqref{e42} can be formulated as the minimization problem of $\mathcal{Q}^{(\tau>0)}(\Lambda,\nu) = \max_{\delta \ge 0}\mathcal{L}^{(\tau>0)}(\boldsymbol{\delta},\Lambda,\nu)$ with $\Lambda \ge 0$ and $\nu$ represent the Lagrangian multipliers related to constraints $\eqref{eq:time_constraint}$ and $\eqref{e42.1}$. Furthermore, for given $\mathbf{P}^{(s-1)}$, $\Lambda$, and $\nu$, with $\Lambda+\nu P_\mathrm{avg}\ge 0 $, the maximizer is \begin{equation}
\hat{\delta}_0^{(\tau>0)}=0, \quad
\hat{\delta}_k^{(\tau>0)}=\frac{\Phi_k P_{\mathrm{avg}}}{q_k^{*}\Phi_k P_k^{(s-1)}}, k\in \mathcal{K},
\label{eq:tau_bar}
\end{equation} where $\Phi_k=\frac{\beta_k H_k(\mathbf{u},\mathbf{z})}{\Upsilon(\tau P_k^{(s-1)}+\sigma^2)}$ and $q_k^{*}$ is the solution of $
\hat{f}(q_k)=\frac{(\Lambda+\nu P_k^{(s-1)})\ln 2}{\alpha_k}$ with $\hat{f}(q)\triangleq f(q) -\frac{\Phi_k P_k^{(s-1)}}{1+q}$ \cite{ju2014optimal1}. Next, we minimize $\mathcal{Q}^{(\tau>0)}(\Lambda,\nu)$ by the ellipsoid method with sub-gradients \begin{align}
 \mu_{\Lambda}&=\sum_{k=0}^{K}\bar{\delta}_k-1,\\
    \mu_{\nu} &=\sum_{k=0}^{K}P_k^{(s-1)}\delta_k^{*(\tau>0)}-P_{\mathrm{avg}}.
\end{align}

After the ellipsoid converges, we can obtain the optimal solution $\Lambda^*$ and $\nu^*$ of the dual function and correspondingly get the optimal solution of problem \eqref{e42}. Therefore, for given $\mathbf{P}^{(s-1)}$, we have $\delta^{(s)}=\delta^{*(\tau>0)}$. Next, denote $\mathbf{P}^{(s)} =[P_0^{(s)},\hat{\mathbf{P}}^{(s)}]$ as the $s$-th iteration with $\hat{\mathbf{P}}^{(s)}=[P_1^{(s)},\ldots,P_K^{(s)}]$. Since for $s \ge 1$, $\tau_0^{(s)}=0$, any $P_0^{(s)}\in[0,P_{\mathrm{peak}}]$ is admissible. Further, update $\hat{\mathbf{P}}^{(s)}$ by gradient projection on \begin{equation}
\mathcal{V}=\Big\{\mathbf{P}:\ \sum_{k=1}^{K}\delta_k^{(s)}P_k=P_{\mathrm{avg}},\ 0\le P_k\le P_{\mathrm{peak}}, k \in \mathcal{K}\Big\}.
\label{eq:feasible_set_E}
\end{equation} 

Specifically, $\hat{\mathbf{P}}^{(s)}$ can be obtained by the gradient projection method, shown as \begin{equation}\hat{\mathbf{P}}^{(s)}=\bar{\mathbf{P}}^{(s-1)}+r^{(s)}\big(\tilde{\mathbf{P}}^{(s)}-\bar{\mathbf{P}}^{(s-1)}\big),
\label{eq:gp_updates}\end{equation} where $\bar{\mathbf{P}}^{(s)}=\mathcal{P}_{\mathcal{V}}\!\Big(\hat{\mathbf{P}}^{(s-1)}+o^{(s)}\mathbf{g}^{(s)}\Big)$ with gradient $\mathbf{g}^{(s)} =[g_1^{(s)},\ldots,g_K^{(s)}]^\top$. Additionally, $o^{(s)}$ is the projected-gradient step size and $r^{(s)}\in(0,1]$ is a relaxation factor. In implementation, both are selected by a simple backtracking line search to ensure ascent of the true WSR objective. Specifically, starting from initial values $o^{(s)}=1$ and $r^{(s)}=1$, they are successively reduced by a factor $\beta_{\mathrm{ls}}\in(0,1)$ until the updated power vector yields a non-decreasing objective value. Finally, let $S = (P_{\mathrm{avg}}-\delta_k^{(s)}P_k^{(s-1)})$, $g_k^{(s)}$ can be given by \begin{equation}
    g_k^{(s)}= -\frac{\alpha_k \delta_k^{(s)}}{\ln 2}
\frac{\beta_k H_k(\mathbf{u},\mathbf{z})\big(\delta_k^{(s)}+\tau S \big)}
{\Upsilon\big(\tau P_k^{(s-1)}+\sigma^2\big)\delta_k^{(s)}+\beta_k H_k(\mathbf{u},\mathbf{z})S}.
\label{eq:gradient_q}
\end{equation}

The proposed alternating procedure yields a non-decreasing WSR sequence. For fixed $P^{(s-1)}$, the time-update subproblem in \eqref{e42} is convex and solved optimally via duality, so the objective cannot decrease after updating $\boldsymbol{\delta}$. For fixed $\boldsymbol{\delta}^{(s)}$, the power vector is updated by projected gradient ascent, and the backtracking line search ensures that each accepted update is also non-decreasing. Therefore, the overall AO procedure generates a monotonically non-decreasing objective sequence. Since the WSR is upper bounded under the feasible constraints, the algorithm converges to a finite limit point. In practice, the algorithm is terminated when $\frac{\big|J^{(s+1)}-J^{(s)}\big|}{\max\{1,|J^{(s)}|\}} \le \varepsilon,
$ where $J^{(s)}$ denotes the WSR value at the $s$-th AO iteration.

The algorithm to solve problem \eqref{e42} for the imperfect SIC case is presented in \textbf{Algorithm \ref{a2}}. In the finite-SI regime, problem \eqref{e42} is solved by AO over the time and power variables. The time update requires $O(K)$ operations, whereas the power update involves $O(K)$ gradient computation and an $O(K^2)$ projection onto $\mathcal{V}$. Hence, the overall per-iteration complexity is dominated by the power-update step and scales as $O(K^2)$.

\begin{algorithm}[t]
\caption{Time-Power Allocation ($\tau>0$)}
\begin{algorithmic}[1] \label{a2}
\STATE \textbf{Input:} $s=0$, $s_{max}=50$, $\varepsilon=10^{-4}$, and $(\boldsymbol{\delta}^{(0)},\mathbf{P}^{(0)})$ from the perfect-SIC solution.
\REPEAT
\STATE \textbf{Time update:} For fixed $\mathbf{P}^{(s)}$, update $\boldsymbol{\delta}^{(s+1)}$ by solving the dual problem in \eqref{e42}.
\STATE \textbf{Power update:} For fixed $\boldsymbol{\delta}^{(s+1)}$, update $\mathbf{P}^{(s+1)}$ via projected gradient ascent with backtracking line search.
\STATE $s\gets s+1$.
\UNTIL {$\frac{|J^{(s)}-J^{(s-1)}|}{\max\{1,|J^{(s-1)}|\}} \le \varepsilon$ or $s \ge s_{\max}$}
\STATE Output: Locally optimal solution $(\boldsymbol{\delta}^{\star},\mathbf{P}^{\star})$.
\end{algorithmic}
\end{algorithm}

\vspace{-1.2em}

\subsection{Antenna Position}

For any given $\{\boldsymbol{\delta}, \mathbf{P}, \mathbf{z}\}$, the subproblem with respect to $\{\mathbf{u}\}$ is given by
\begin{align} \label{e52}
\max_{\mathbf{u}}\; &\sum^K_{k=1}\alpha_kR_k(\mathbf{u}) \\
\textrm{s.t.} \quad & \eqref{eq:position_constraint}, \eqref{eq:separation_constraint}. \notag
\end{align}

The achievable rate in this section is written as
\begin{equation}
R_k(\mathbf{u}) =\delta_k \log_2\!\big(1+\kappa_k\,\mathbf{H}_k(\mathbf{u},\mathbf{z})\big),
\end{equation} \begin{equation} \kappa_k \triangleq \frac{\beta_k}{\Upsilon(\tau P_k+\sigma^2)}\cdot \frac{1}{\delta_k}\sum_{i\neq k}\delta_i P_i.
\label{eq:Ri_kappa}
\end{equation}

Considering the composite channel gain in \eqref{e21} and \eqref{e22}, we define $\mathbf{P}_{RX}\triangleq \mathbf{S}_{RX}^\top \mathbf{S}_{RX}$ and $\mathbf{P}_{TX}\triangleq \mathbf{S}_{TX}^\top \mathbf{S}_{TX}$ as $M\times M$ diagonal projection matrices to ensure dimensional consistency in the gradient expressions below. For notational brevity, we suppress the explicit dependence on $\mathbf{z}$ and simply write $\mathbf{H}_k(\mathbf{u})$, $\mathbf{G}_{D,k}(\mathbf{u})$, and $\mathbf{G}_{U,k}(\mathbf{u})$ in this subsection. Further, we denote $\mathcal{J}(\mathbf{u})=\sum_{k=1}^K \alpha_k R_k(\boldsymbol{\delta},\mathbf{P},\mathbf{u},\mathbf{z})$ as the WSR objective.

Furthermore, we derive the first-order information. By the chain and product rules, we have \begin{equation}
\frac{\partial R_k}{\partial \mathbf{H}_k(\mathbf{u})}=\frac{\delta_k}{\ln 2}\cdot \frac{\kappa_k}{1+\kappa_k \mathbf{H}_k(\mathbf{u})},
\end{equation} \begin{equation}
\nabla_{\mathbf{u}} \mathbf{H}_k(\mathbf{u})
=\mathbf{G}_{D,k}(\mathbf{u})\,\nabla_{\mathbf{u}} \mathbf{G}_{U,k}(\mathbf{u})+\mathbf{G}_{U,k}(\mathbf{u})\,\nabla_{\mathbf{u}} \mathbf{G}_{D,k}(\mathbf{u}).
\end{equation}

Combining the selection projections, the gain gradients with respect to a single element position component $\mathbf{u}_m\in[x_m,z_m]$ are \begin{align}
\frac{\partial \mathbf{G}_{U,k}(\mathbf{u})}{\partial \mathbf{u}_m}
&=2\,\Re\!\left\{\Big(\frac{\partial \mathbf{h}_k^{\mathrm{UL}}}{\partial \mathbf{u}_m}\Big)^{\!H} \mathbf{P}_{\!RX}\,\mathbf{h}_k^{\mathrm{UL}}\right\},\\
\frac{\partial \mathbf{G}_{D,k}(\mathbf{u})}{\partial \mathbf{u}_m}
&=2\,\Re\!\left\{\Big(\frac{\partial \mathbf{h}_k^{\mathrm{DL}}}{\partial \mathbf{u}_m}\Big)^{\!H} \mathbf{P}_{\!TX}\,\mathbf{h}_k^{\mathrm{DL}}\right\}.
\end{align}

To retain the geometric NF spherical-wave characteristics captured by the adopted channel model, the element-user response is factorized as $h_m(\mathbf{u})=a(r_m)e^{j\psi_m(\mathbf{u})}$,  where $a(r)\propto r^{-\nu}$, $r_m$ is the geometric distance, and $\mathbf{\psi}_m(\mathbf{u})$ is the spherical-wave phase from the array element to the user. Thus, using $\partial r_m/\partial x_m=(x_m-x_{\mathrm{u}})/r_m$ and $\partial r_m/\partial z_m=(z_m-z_{\mathrm{u}})/r_m$, we obtain \begin{align}
\frac{\partial h_m}{\partial u_m}
&=h_m\!\left(\frac{\partial \ln a}{\partial u_m}+j\,\frac{\partial \psi_m}{\partial u_m}\right),\\
\frac{\partial \ln a}{\partial u_m}
&=-\nu\,\frac{1}{r_m}\frac{\partial r_m}{\partial u_m},
\end{align} thereby explicitly incorporating both the distance-dependent amplitude variation and the spherical-wave phase variation with respect to the antenna position into the closed-form computation of $\nabla_{\mathbf{u}}\mathcal{J}$.

We next consider constraints and feasibility. The position vector must satisfy the per-FA box domain $\mathbf{u}_m\in\mathcal{C}_m, m\in \mathcal{M}$ and the minimum spacing $\| \mathbf{u}_m - \mathbf{u}_i \|\ge d_{\mathrm{min}}, i\in \mathcal{M}, i \neq m$ for any two elements. Since the spacing set is nonconvex, we instead consider the modified objective \begin{equation}
\widetilde{\mathcal{J}}(\mathbf{u})
=\mathcal{J}(\mathbf{u})-\lambda_c\sum_{m<i}\frac{1}{\max\!\big(\varepsilon,\ \| \mathbf{u}_m - \mathbf{u}_i \|-d_{\min}\big)},
\label{e58}
\end{equation} where $\lambda_c>0$ is a penalty weight and $\varepsilon$ is a small regularization constant introduced to avoid numerical instability when $\| \mathbf{u}_m - \mathbf{u}_i \|$ approaches $d_{\min}$. The additional term serves as a regularized collision-avoidance penalty to approximately enforce the minimum-spacing constraint within the MM/SCA-based position update. In this way, only Euclidean projection onto $\mathcal{C}\triangleq\prod_m \mathcal{C}_m$ is required, which keeps the first-order update implementable and numerically stable. In practice, $\lambda_c$ is chosen sufficiently large, while $\varepsilon$ is set as a small safeguard parameter. Alternative treatments based on slack variables or exact projection are possible, but would lead to a more involved inner feasibility update.

Additionally, we present two inner solvers consistent with the MM/SCA-based method. At the current iterate $\mathbf{u}^{(t)}$, lower-bound the concave $\log(1+x) \mathbf{H}_k(\mathbf{u})$ at $\mathbf{u}^{(t)}$, and subtract a conservative quadratic term to form a strongly concave global lower bound \begin{align} \label{e60}
g^{(t)}(\mathbf{u}) = \sum_{k=1}^K \alpha_k\!\left[c_k^{(t)} 
+b_k^{(t)}\mathcal{H}(\mathbf{u})^{(t)}\!\right], 
\end{align} where \begin{align} 
\mathcal{H}(\mathbf{u})^{(t)} 
&= \mathbf{H}_k(\mathbf{u}^{(t)})+\nabla \mathbf{H}_k(\mathbf{u}^{(t)})^\top(\mathbf{u}-\mathbf{u}^{(t)})-\frac{\ell_k}{2}\|\mathbf{u}-\mathbf{u}^{(t)}\|^2,\\
c_k^{(t)} 
&=\delta_k\log_2\!\big(1+x_k^{(t)}\big)-\frac{\delta_k}{\ln 2}\frac{x_k^{(t)}}{1+x_k^{(t)}},\\
b_k^{(t)}
&=\frac{\delta_k}{\ln 2}\frac{\kappa_k}{1+x_k^{(t)}},\\ 
x_k^{(t)}
&=\kappa_k \mathbf{H}_k(\mathbf{u}^{(t)}),
\end{align} and $\ell_k$ is chosen as an upper bound on the spectral norm of the Hessian of $\nabla^2 \mathbf{H}_k(\mathbf{u})$ (using empirical or relaxed bounds). Finally, maximizing $g^{(t)}(\mathbf{u})$ yields a Newton-like direction in the unconstrained case; with backtracking line search and box projection, the update is \begin{equation}
\mathbf{u}^{(t+1)}=\Pi_{\mathcal{C}}\!\big(\mathbf{u}^{(t)}+\eta_t\,\mathbf{d}^{(t)}\big),
\label{eq:mm_update}
\end{equation} where $\mathbf{d}^{(t)}$ is the Newton-like ascent direction given by \eqref{e60}, and $\eta_t$ is chosen by Armijo backtracking. Under standard MM conditions, the weighted sum-rate is monotonically nondecreasing and the sequence converges to a stationary value. For the MM/SCA for FAs positioning, the per-iteration complexity is $\mathcal{O}((1+L_{\text{ls}})MK)$ and the $T$-iteration runtime is $\mathcal{O}(T(1+L_{\text{ls}})MK)$, with memory $\mathcal{O}(M+K)$. The MM/SCA-based algorithm is shown in \textbf{Algorithm \ref{a3}}.

\begin{algorithm}[t]
\caption{FA Position Optimization}\label{a3}
\begin{algorithmic}[1]
\STATE \textbf{Input:} initial feasible FA position $\mathbf{u}^{(0)}$, $\lambda_c = 10$, $\varepsilon =10^{-4}$, $T_{\max} =30$, and $\{\boldsymbol{\delta},\mathbf{P},\mathbf{z}\}$.
\STATE Set iteration index $t \leftarrow 0$.
\REPEAT
\STATE Compute $H_k(\mathbf{u}^{(t)},z)$ and $J(\mathbf{u}^{(t)})$.
\STATE Form the regularized collision-avoidance objective $\tilde{J}(\mathbf{u})$.
\STATE Compute $\nabla \tilde{J}(\mathbf{u}^{(t)})$.
\STATE Construct a local quadratic surrogate $\tilde{g}^{(t)}(\mathbf{u})$.
\STATE Update $\mathbf{u}^{(t+1)}$ by projected ascent over $\mathcal{C}$.
\STATE $t \leftarrow t+1$.
\UNTIL{$\frac{|J(\mathbf{u}^{(t)})-J(\mathbf{u}^{(t-1)})|}{\max\{1,|J(\mathbf{u}^{(t-1)})|\}} \le \varepsilon$ or $t \ge T_{\max}$}
\STATE \textbf{Output:} $\mathbf{u}^{\star} \leftarrow \mathbf{u}^{(t)}$.
\end{algorithmic}
\end{algorithm}
\vspace{-0.3\baselineskip}

\subsection{Antenna Group Selection}

For any given $\{\boldsymbol{\delta}, \mathbf{P}, \mathbf{u}\}$, the subproblem with respect to $\{\mathbf{z}\}$ is given by \begin{align} \label{e67}
&\max_{\mathbf{z}}\
\sum^K_{k=1}\alpha_kR_k(\mathbf{z}) \\
&\textrm{s.t.}\
\eqref{eq:group_constraint}, \eqref{eq:var_range_discrete}. \notag
\end{align} 

Based on \eqref{eq:Ri_kappa}, the achievable rate can be expressed as \begin{equation}
R_k(\mathbf{z}) =\delta_k \log_2\!\big(1+\kappa_k\,\mathbf{H}_k(\mathbf{u},\mathbf{z})\big).
\end{equation}

For notational brevity, we suppress the explicit dependence on $\mathbf{u}$ and simply write $\mathbf{H}_k(\mathbf{z})$, $\mathbf{G}_{D,k}(\mathbf{z})$, and $\mathbf{G}_{U,k}(\mathbf{z})$ in this subsection. 

Let $g(m)$ denote the group index of element $m$, i.e.,
$m\in\mathcal G_{g(m)}$. From \eqref{e21} and \eqref{e22}, for
binary $\mathbf{z}$ we can equivalently express
\begin{align}
    \mathbf{G}_{U,k}(\mathbf{z})
    &= \sum_{m=1}^M \bigl(1 - z_{g(m)}\bigr)
       \bigl|h^{\mathrm{UL}}_{k,m}(\mathbf{u})\bigr|^2, \label{eq:GU-linear}\\
    \mathbf{G}_{D,k}(\mathbf{z})
    &= \sum_{m=1}^M z_{g(m)}
       \bigl|h^{\mathrm{DL}}_{k,m}(\mathbf{u})\bigr|^2. \label{eq:GD-linear}
\end{align}
To obtain a differentiable relaxation over the convex set
$\mathcal{Z} \triangleq \{\mathbf{z}\in[0,1]^N:\mathbf{1}^\top\mathbf{z}=N/2\}$,
we replace the binary gates by smooth quadratic gates, leading to \begin{align}
    \mathbf{G}_{U,k}(\mathbf{z}) &=
       \sum_{m=1}^M \bigl(1 - z_{g(m)}\bigr)^2
       \bigl|h^{\mathrm{UL}}_{k,m}(\mathbf{u})\bigr|^2, \label{eq:GU-relaxed}\\
    \mathbf{G}_{D,k}(\mathbf{z}) &=
       \sum_{m=1}^M z_{g(m)}^{2}
       \bigl|h^{\mathrm{DL}}_{k,m}(\mathbf{u})\bigr|^2. \label{eq:GD-relaxed}
\end{align}

Define $x_k(\mathbf{z})=\kappa_k \mathbf{H}_k(\mathbf{z})$ and $x_k^{(t)}=\kappa_k \mathbf{H}_k(\mathbf{z}^{(t)})$. Since $\log(1+x)$ is concave and $C^1$ on $x\!\ge\!0$, a quadratic global lower bound holds \begin{align} \label{e70}
&\log\!\big(1+x_k(\mathbf{z})\big)\ge \notag \\ & \quad \log(1+x_k^{(t)})+\frac{x_k(\mathbf{z})-x_k^{(t)}}{1+x_k^{(t)}}-\frac{1}{2}\big(x_k(\mathbf{z})-x_k^{(t)}\big)^2.
\end{align}

To make the right-hand side concave in $\mathbf{z}$, minorize $\mathbf{H}_k(\mathbf{z})$ by a concave quadratic via the descent lemma: for any $L_{H,k}$ not smaller than the Lipschitz constant of $\nabla \mathbf{H}_k(\mathbf{z})$,
\begin{equation}\label{eq:H_minor}
\mathbf{H}_k(\mathbf{z})\;\ge\;\mathbf{H}_k^{(t)}+\nabla \mathbf{H}_k^{(t)\top}(\mathbf{z}-\mathbf{z}^{(t)})-Z
\;\triangleq\;\mathbf{H}_k^{(t)}(\mathbf{z}),
\end{equation} where $Z = \frac{L_{H,k}}{2}\|\mathbf{z}-\mathbf{z}^{(t)}\|^2$. Substituting $x_k(\mathbf{z})=\kappa_k \mathbf{H}_k(\mathbf{z})$ into \eqref{e70} and replacing $\mathbf{H}_k(\mathbf{z})$ with $\mathbf{H}_k^{(t)}$ yield \begin{align}\label{e72}
    &\log\!\big(1+x_k(\mathbf{z})\big)\ge C_k^{(t)} + \notag \\ & \quad\alpha_k^{(t)}\Big(\nabla \mathbf{H}_k^{(t)\top}(\mathbf{z} -\mathbf{z}^{(t)})-Z\Big) -\beta_k^{(t)}\|\mathbf{z}-\mathbf{z}^{(t)}\|^2,
\end{align}where \begin{align}
\alpha_k^{(t)}
&=\frac{\kappa_k}{1+\kappa_k \mathbf{H}_k^{(t)}},\\
\beta_k^{(t)}
&=\frac{\kappa_k^2 L_{H,k}^2}{2\,(1+\kappa_k \mathbf{H}_k^{(t)})^2},\\
C_k^{(t)}
&=\log(1+\kappa_k \mathbf{H}_k^{(t)})-\frac{\kappa_k \mathbf{H}_k^{(t)}}{1+\kappa_k \mathbf{H}_k^{(t)}}.
\end{align}

Then, summing \eqref{e72} with weights $\alpha_k\delta_k$ gives the iteration $t$ surrogate \begin{align}\label{eq:surrogate}
&g^{(t)}(\mathbf{z})= \sum_{k=1}^{K}\alpha_k\delta_k \notag \\ & \quad\Big[
C_k^{(t)}+\alpha_k^{(t)}\,\nabla \mathbf{H}_k^{(t)\top}(\mathbf{z}-\mathbf{z}^{(t)})-\tfrac{1}{2}\,\Gamma_k^{(t)}\|\mathbf{z}-\mathbf{z}^{(t)}\|^2\Big],
\end{align} where $\Gamma_k^{(t)}\triangleq \alpha_k^{(t)}L_{H,k}+2\beta_k^{(t)}$. By construction, $g^{(t)}(\mathbf{z})\le \mathcal{J}(\mathbf{z})$ for all $\mathbf{z}\in\mathcal{Z}$ and $g^{(t)}(\mathbf{z}^{(t)})=\mathcal{J}(\mathbf{z}^{(t)})$. Next, maximizing \eqref{eq:surrogate} over $\mathcal{Z}$ is a concave quadratic problem, and the objective function of the problem is formulated as \begin{equation}\label{eq:qp}
\max_{\mathbf{z}\in\mathcal{Z}}\;\; b^{(t)\top}\mathbf{z}-\tfrac{1}{2}\mathbf{z}^\top \mathbf{Q}^{(t)}\mathbf{z},\end{equation} where $b^{(t)}=\sum_{k}\alpha_k\delta_k\,\alpha_k^{(t)}\,\nabla \mathbf{H}_k^{(t)}+\mathbf{Q}^{(t)}\mathbf{z}^{(t)}$, and $\mathbf{Q}^{(t)}=\Big(\sum_{k}\alpha_k\delta_k\,\Gamma_k^{(t)}\Big) \mathbf{I}$. Since $\mathbf{Q}^{(t)}$ is a scaled identity matrix, the projected ascent method can be used with an exact projection onto $\mathcal{Z}$. Letting $\mathbf{z}^{(t+1)}$ be the maximizer, an optional line search on the segment $\mathbf{z}^{(t)}+\eta\big(\mathbf{z}^{(t+1)}-\mathbf{z}^{(t)}\big)$ with $\eta\in(0,1]$ ensures monotone ascent of the true objective.

 \begin{algorithm}[t]
\caption{Antenna Group Selection}
\label{a4}
\begin{algorithmic}[1]
\STATE \textbf{Input:} $\{\boldsymbol{\delta},\mathbf{P},\mathbf{u}\}$, initial binary grouping $\mathbf{z}^{(0)}$, relaxed set $\mathcal{Z}$, $T_{\max}=30$, $\varepsilon = 10^{-4}$, $\eta=1$.
\STATE Relax $\mathbf{z}^{(0)}$ to $\mathcal{Z} = \{\mathbf{z}\in[0,1]^N : \mathbf{1}^\top \mathbf{z}=N/2\}$ and set $t \leftarrow 0$.
\REPEAT
    \STATE Compute $\mathbf{G}_{U,k}(\mathbf{z}^{(t)})$, $\mathbf{G}_{D,k}(\mathbf{z}^{(t)})$, composite gains $\mathbf{H}_k(\mathbf{z}^{(t)})$ and WSR $\mathcal{J}(\mathbf{z}^{(t)})$.
    \STATE Build a concave quadratic surrogate $g^{(t)}(\mathbf{z})$ of $\mathcal{J}(\mathbf{z})$ on $\mathcal{Z}$.
    \STATE Update $\mathbf{z}^{(t+1)}$ by maximizing $g^{(t)}(\mathbf{z})$ over $\mathcal{Z}$ using projected gradient/ascent.
    \STATE Set $t \leftarrow t+1$.
\UNTIL{$|\mathcal{J}(\mathbf{z}^{(t)})-\mathcal{J}(\mathbf{z}^{(t-1)})| \le \varepsilon$ or $t \ge T_{\max}$}
\STATE Quantize $\mathbf{z}^{(t)}$ to a binary vector with $N/2$ ones by keeping the largest $N/2$ entries, and denote the result by $\mathbf{z}^*$.
\STATE \textbf{Output:} binary group selection vector $\mathbf{z}^*$.
\end{algorithmic}
\end{algorithm}

\paragraph*{Remark} The gradient $\nabla \mathbf{H}_k(\mathbf{z})=\mathbf{G}_{D,k}(\mathbf{z})\nabla \mathbf{G}_{U,k}(\mathbf{z})+\mathbf{G}_{U,k}(\mathbf{z})\nabla \mathbf{G}_{D,k}(\mathbf{z})$ is separable over groups, hence both $\nabla \mathbf{H}_k^{(t)}$ and crude $L_{H,k}$ are easy to compute. Initialize $\mathbf z^{(0)}$ by ranking groups according to the channel-gain metric $s_n=\sum_{k=1}^{K}\left(\sum_{m\in G_n}|h^{\mathrm{DL}}_{k,m}(\mathbf u)|^2-\sum_{m\in G_n}|h^{\mathrm{UL}}_{k,m}(\mathbf u)|^2\right), n\in\mathcal N$, and setting the groups corresponding to the largest $N/2$ values of $\{s_n\}$ to one. After SCA converges in $\mathcal{Z}$, threshold the largest $N/2$ entries to $1$ and the rest to $0$, then apply a few pairwise swaps that only accept improvements of the true WSR. If a penalty $-\lambda_{\mathrm{SI}}\|\mathbf{H}_{\mathrm{SI}}(\mathbf{z})\|_F^2$ is present, note it is concave (negative of a convex norm-square in linear gates) and can be retained in \eqref{eq:surrogate} without breaking concavity.

After solving the relaxed problem over $\mathcal Z$, the final binary grouping is obtained by setting the largest $N/2$ entries to one and the remaining entries to zero. Starting from this thresholded point, a pairwise-swap local refinement is further performed, where one TX group and one RX group are exchanged and the swap is accepted only if it improves the true WSR objective. The procedure stops when no improving swap exists. The algorithm of solving problem \eqref{e67} is shown in \textbf{Algorithm \ref{a4}}, where each iteration costs $O(NK)$ to form $b^{(t)}$ from cached per-user terms plus $O(N\log N)$ for a single projection onto $\mathcal{Z}$. 

\section{Numerical Results}\label{s5}

In this section, numerical results are presented to verify the effectiveness of the proposed FD-FANS scheme. In particular, we evaluate the performance of the proposed design from multiple perspectives, including the ASR, EE, Jain's fairness index, computational complexity, convergence behavior, and robustness against residual SI as well as imperfect CSI. To provide a comprehensive performance comparison, the proposed FD-FANS is benchmarked against several representative schemes, including (i) the HD-FANS; 
(ii) the FD-FPANS, which serves as the fixed-antenna FD benchmark under the same residual-SI model; 
(iii) the non-grouped FD-FANS; 
(iv) a far-field counterpart under the same optimization framework; and 
(v) a fixed asymmetric TX/RX partition strategy with \(1/3\) TX groups and \(2/3\) RX groups. In addition, imperfect-CSI cases with different uncertainty factors are included to further examine the robustness of the proposed design under practical channel uncertainty.


\subsection{Simulation Setup}

We consider a BS UPA with $M=64$ antennas, arranged with half-wavelength element spacing and partitioned into $N=4$ disjoint groups. The system operates in the mmWave band with carrier frequency $f_c=90$ GHz. For simplicity, the movable region of each FA is modeled as a square. Accordingly, the bounded area of FA along the $x$- or $z$-axis is denoted by $\sqrt{\mathcal{C}_m} = 0.05$ m. Large-scale propagation uses equal downlink and uplink path-loss exponents. The constant factor $10^{-3}$ corresponds to a $30$ dB average path loss at the $1$ m reference distance. Modulation loss is represented by a standard SINR gap $\Upsilon=9.8$ dB in the rate expression. For the FD operation, the post-analog-cancellation loopback component has a residual power ratio $\phi$ that we report results for the ideal case (perfect SIC) and for representative SI levels $\phi\in\{-80,-60,-40\}$ dB. The values of the quantization error coefficient $\zeta=-60$ dB, the channel estimation error coefficient $\epsilon =-60$ dB, and the combined energy conversion and transmission efficiency $\beta_k= \gamma_k \eta_k=0.5^8$ are adopted based on the parameter settings reported in \cite{ju2014optimal}. In addition, to examine robustness against channel uncertainty, we additionally evaluate the proposed FD-FANS under imperfect CSI with uncertainty factors \(\xi \in \{0.05,\,0.1\}\). 

Unless otherwise specified, all results are averaged over $50$ independent Monte Carlo realizations. The performance ordering is consistent across realizations, and the variation is small relative to the performance gaps. For clarity of presentation, confidence intervals are not explicitly shown in the figures, as they are relatively narrow and do not affect the observed performance trends. Simulation parameters for the proposed system are summarized in Table \ref{table1}.

\captionsetup[table]{
  labelsep=newline,
  textfont=sc}

\begin{table}[t]
\centering
\caption{System Parameters}
\label{table1}
\resizebox{\linewidth}{!}{
\begin{tabular}{lc}
\hline
\textbf{Parameter} & \textbf{Value} \\
\hline
Number of antennas at the BS & $M = 64$ \\
Number of antenna groups & $N = 4$ \\
Carrier frequency & $f_{c} = 90$ GHz \\
Wavelength & $\lambda = 0.003$ m \\
Distance between user $k$ and the BS & $d_{k} \in [5,10]$ m \\
Azimuth angle range of user $k$ to the BS & $\theta_{k} \in [30^{\circ},120^{\circ}]$ \\
Elevation angle range of user $k$ to the BS & $\varphi_{k} \in [30^{\circ},120^{\circ}]$ \\
Movable area for each FA & $\mathcal{C}_m = 0.0025~\text{m}^2$ \\
Length of coherent communication time block & $T = 1$ \\
Path loss at $1$ m & $\rho_k = 30$ dB \\
Average noise power for users & $\sigma_u^2 = -50$ dBm \\
SINR gap & $\Upsilon = 9.8$ dB \\
Quantization error coefficient & $\zeta = -60$ dB \\
Channel estimation error coefficient & $\epsilon = -60$ dB \\
Combined energy conversion and transmission efficiency & $\beta_k = \gamma_k \eta_k = 0.5^8$ \\
\hline
\end{tabular}}
\end{table}

\subsection{Complexity Discussion}

\captionsetup[table]{
  labelsep=newline,
  justification=centering,
  singlelinecheck=false,
  textfont=sc}
  
\begin{table}[t]
\centering
\caption{Computational complexity comparison of the proposed method and baseline schemes}
\label{tab:complexity_comparison}
\resizebox{\linewidth}{!}{
\begin{tabular}{lcc}
\hline
\textbf{Scheme} & \textbf{Optimized Variables} & \textbf{Per-iteration Complexity} \\
\hline
Proposed FD-FANS & $\{\boldsymbol{\delta},\mathbf{P},\mathbf{u},\mathbf{z}\}$ 
& $\mathcal{O}\!\left(K^{2}+MK+NK\right)$ \\

HD-FANS & $\{\boldsymbol{\delta},\mathbf{P},\mathbf{u}\}$ 
& $\mathcal{O}\!\left(K+MK\right)$ \\

FD-FPANS & $\{\boldsymbol{\delta},\mathbf{P},\mathbf{z}\}$ 
& $\mathcal{O}\!\left(K^{2}+NK\right)$ \\

Non-grouped FD-FANS & $\{\boldsymbol{\delta},\mathbf{P},\mathbf{u}\}$ 
& $\mathcal{O}\!\left(K^{2}+MK\right)$ \\
\hline
\end{tabular}}
\end{table}

Table~\ref{tab:complexity_comparison} summarizes the computational complexity of the proposed AO-based FD-FANS algorithm and the considered baseline schemes. The overall computational complexity of the proposed AO-based algorithm is determined by the iterative updates of time–power allocation, antenna positions, and grouping variables. Specifically, the per-iteration complexity is dominated by the position optimization and grouping update steps, leading to a polynomial complexity in terms of the number of antennas and users. In comparison, the baseline schemes involve simplified designs without joint optimization over antenna positions and grouping, and thus exhibit lower computational complexity. However, these schemes cannot fully exploit the additional spatial degrees of freedom provided by antenna mobility and grouping. Therefore, the proposed algorithm achieves a favorable tradeoff between computational complexity and performance gain, making it suitable for practical implementation with moderate computational resources.

\subsection{Simulation Results}

\begin{figure}[t]
    \centering
    \begin{subfigure}[t]{0.4\textwidth}
        \centering
        \includegraphics[width=\textwidth]{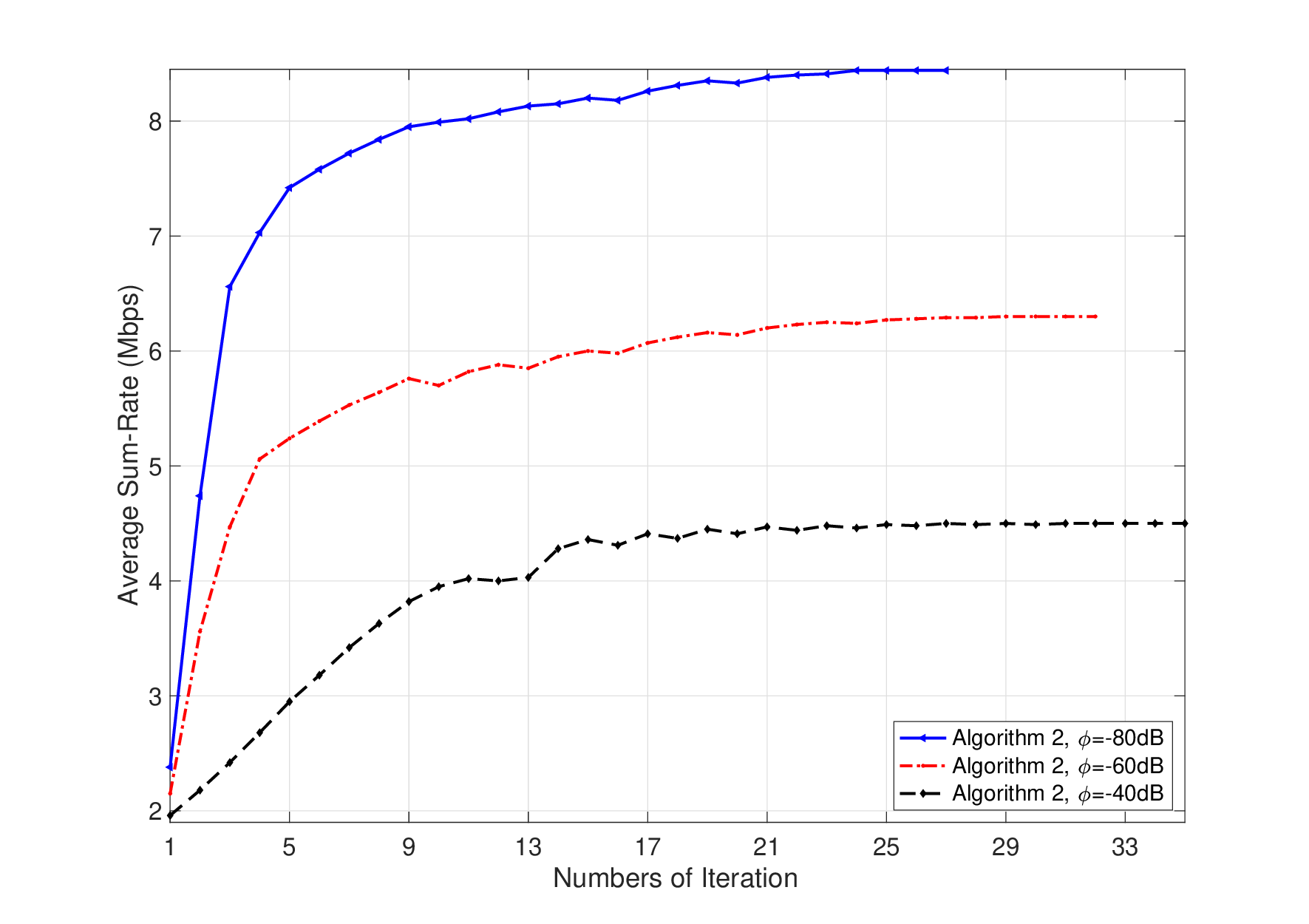}
        \caption{Convergence behavior of Algorithm~2 under different residual SI levels.}
        \label{f1.42}
    \end{subfigure}
    \hfill
    \begin{subfigure}[t]{0.4\textwidth}
        \centering
        \includegraphics[width=\textwidth]{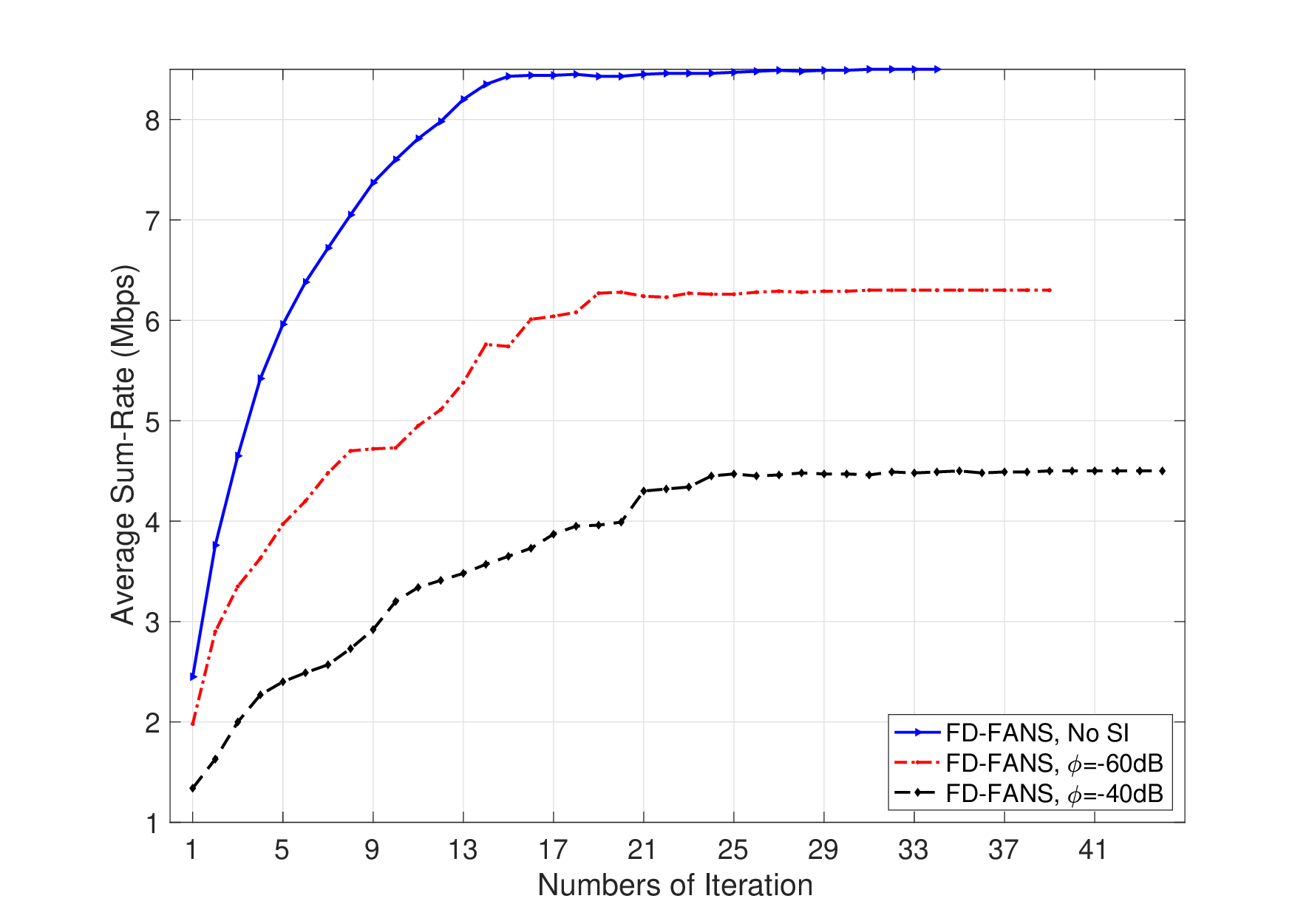}
        \caption{Convergence behavior of the overall AO algorithm under different residual SI levels.}
        \label{f1.41}
    \end{subfigure}
    \caption{Convergence performance of the proposed iterative framework.}
    \label{f1.4}
        \vspace{-2em}
\end{figure}

Fig. \ref{f1.4} illustrates the convergence performance of the proposed iterative framework. As shown in Fig. \ref{f1.42}, it can be observed that \textbf{Algorithm~2} converges reliably under different SI conditions, which confirms the numerical stability of the proposed time-power allocation update in the practical case with finite SI. Also, Fig. \ref{f1.41} further shows that the overall AO algorithm exhibits a clear convergence trend under all considered residual SI levels. The objective value increases rapidly in the initial iterations and then gradually approaches a stable region, indicating that the proposed alternating optimization framework can effectively coordinate the updates of time-power allocation, antenna position optimization, and group selection. Moreover, different residual SI levels mainly affect the final converged performance, while the overall convergence behavior remains stable. Overall, these results verify that the proposed iterative design is computationally well-behaved and can reach stable solutions within a reasonable number of iterations.

\begin{figure}[t]
    \centering
    \includegraphics[width=90mm]{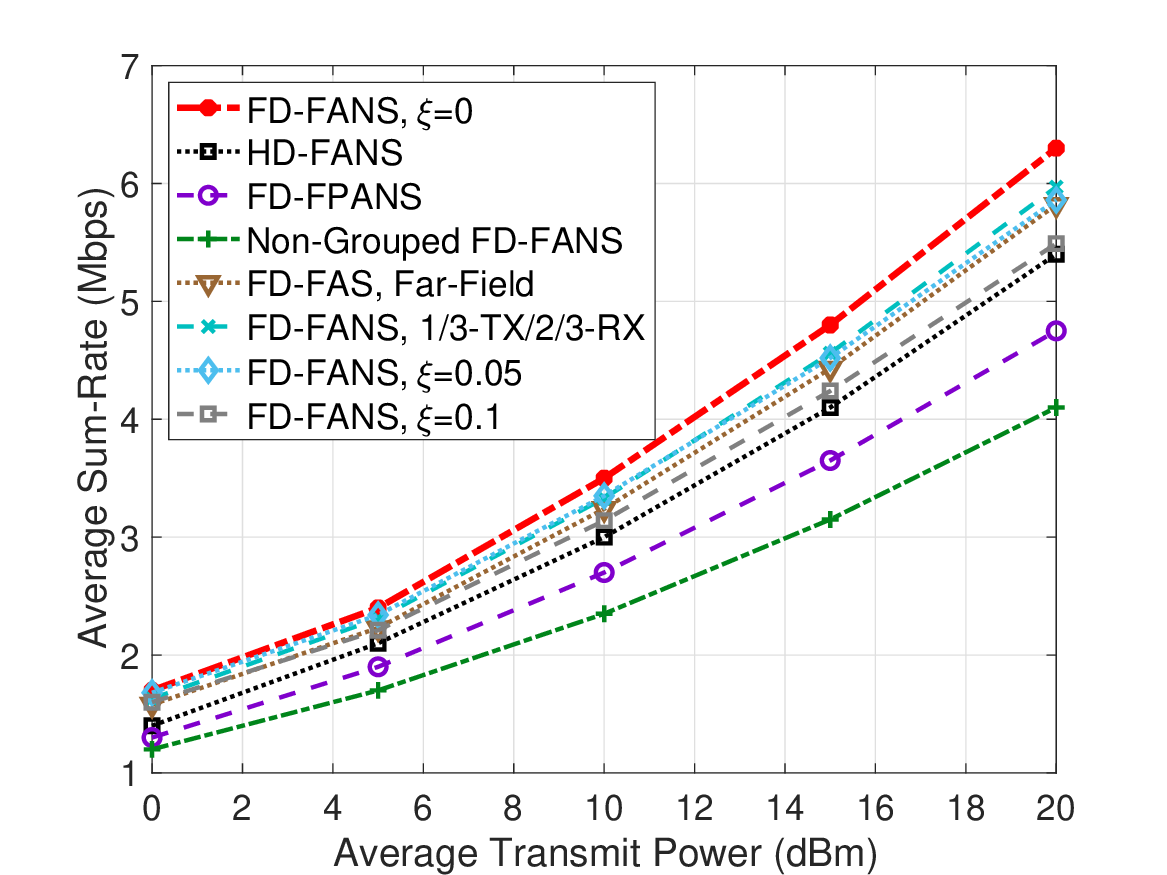}
    \caption{ASR versus $P_{\mathrm{avg}}$ with different CSI uncertainty factors \(\xi\), $P_\mathrm{peak} = 2P_\mathrm{avg}$, $K =10$, and $\phi = -60$ dB}
    \label{f1}
            \vspace{-1em}
\end{figure}

Fig. \ref{f1} plots the ASR versus the ATP \(P_{\mathrm{avg}}\) with \(P_{\mathrm{peak}}=2P_{\mathrm{avg}}\), different CSI uncertainty factors \(\xi\), \(K=10\), and \(\phi=-60\) dB. As expected, the ASR increases monotonically with \(P_{\mathrm{avg}}\), while the growth rate gradually slows down due to the logarithmic rate law and the increasing impact of residual SI at high transmit power. Across the whole range, the proposed FD-FANS consistently achieves the best performance. Compared with the far-field counterpart, the proposed FD-FANS achieves an additional gain of about \(0.2\) Mbps at \(P_{\mathrm{avg}}=10\) dBm and about \(0.5\) Mbps at \(P_{\mathrm{avg}}=20\) dBm. Moreover, the proposed adaptive grouping design also provides a consistent advantage over the baseline with asymmetric partitions with gains of roughly \(0.2\) Mbps at \(P_{\mathrm{avg}}=10\) dBm and \(0.3\) Mbps at \(P_{\mathrm{avg}}=20\) dBm, indicating that a non-adaptive asymmetric partition cannot fully exploit the TX/RX tradeoff. The advantage widens in the medium-to-high power regime, indicating that joint position control and group selection both preserve NF focusing gains and mitigate SI coupling as power grows. Also, introducing CSI uncertainty leads to a gradual reduction in the ASR of the proposed FD-FANS, and a larger uncertainty factor $\xi$ results in more pronounced performance loss. Nevertheless, the proposed scheme still preserves a clear rate improvement as $P_{\mathrm{avg}}$ increases and remains competitive under imperfect CSI, indicating a certain degree of robustness against channel uncertainty.

\begin{figure}[t]
    \centering
    \includegraphics[width=90mm]{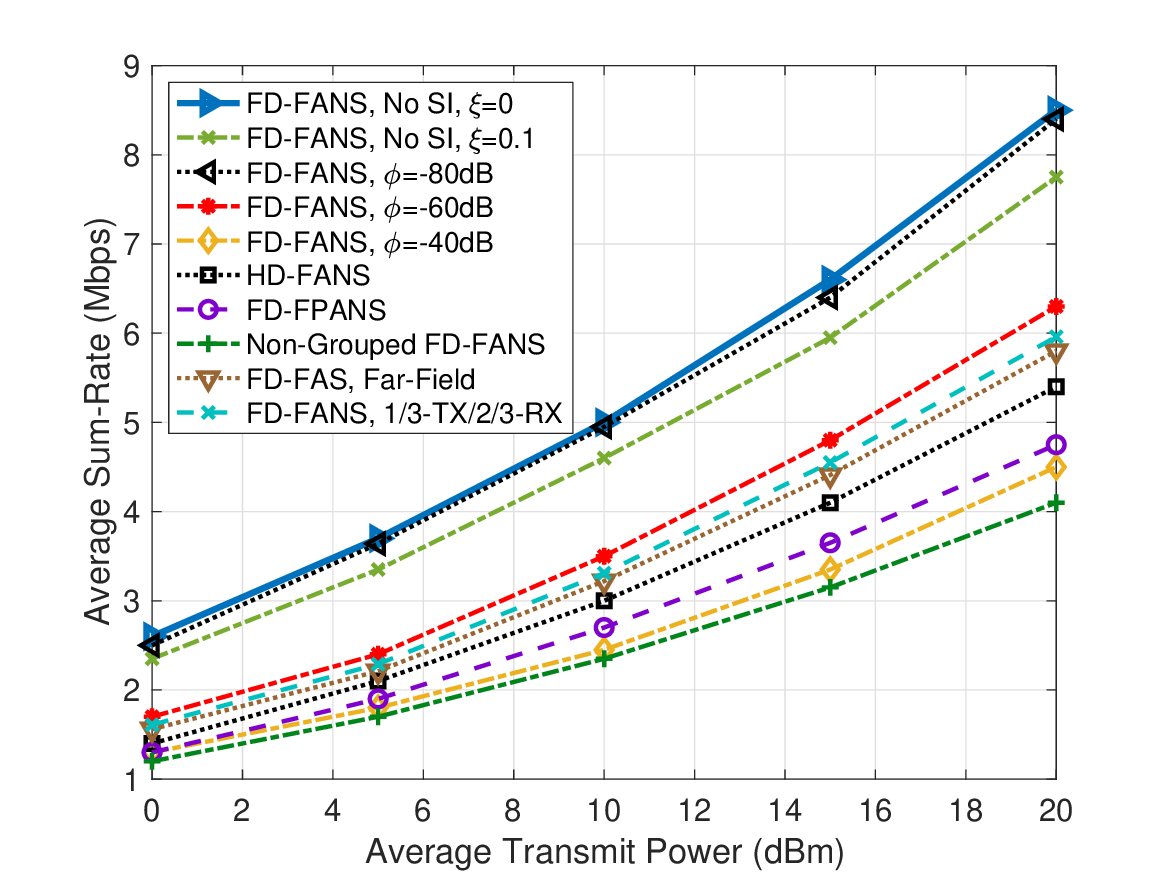}
    \caption{ASR versus $P_\mathrm{avg}$ with different values of $\phi$, $P_\mathrm{peak} = 2P_\mathrm{avg}$, and $K = 10$}
    \label{f2}
            \vspace{-1em}
\end{figure}

Fig.~\ref{f2} shows ASR versus the average power $P_{\mathrm{avg}}$ for different residual SI levels $\phi$, with $P_{\mathrm{peak}}=2P_{\mathrm{avg}}$ and $K=10$. Under perfect SIC, introducing CSI uncertainty with $\xi=0.1$ causes a moderate ASR reduction across all $P_{\mathrm{avg}}$, while the overall increasing trend remains unchanged. Performance degrades predictably as $\phi$ worsens, with a modest loss at low power that becomes pronounced at high power because residual SI scales with $P_{\mathrm{avg}}$ and can dominate the noise. Despite this, FD-FANS remains superior to all benchmarks across $\phi$, demonstrating the robustness of the AO design, where time–power allocation and TX/RX group selection adaptively stagger energy broadcasting to protect uplink decoding under strong SI. The ordering is consistent across powers, with no SI $>-80$ dB$>-60$ dB$>-40$ dB. For the representative case \(\phi=-60\) dB, the proposed FD-FANS remains about \(0.3\) Mbps higher than the far-field baseline at \(P_{\mathrm{avg}}=10\) dBm and about \(0.5\) Mbps higher at \(P_{\mathrm{avg}}=20\) dBm, showing that the NF gain is preserved under residual SI. It also outperforms the baseline with asymmetric partitions by roughly \(0.2\) Mbps at \(P_{\mathrm{avg}}=10\) dBm and about \(0.3\) Mbps at \(P_{\mathrm{avg}}=20\) dBm, which confirms the benefit of adaptive grouping.

\begin{figure}[t]
    \centering
    \includegraphics[width=90mm]{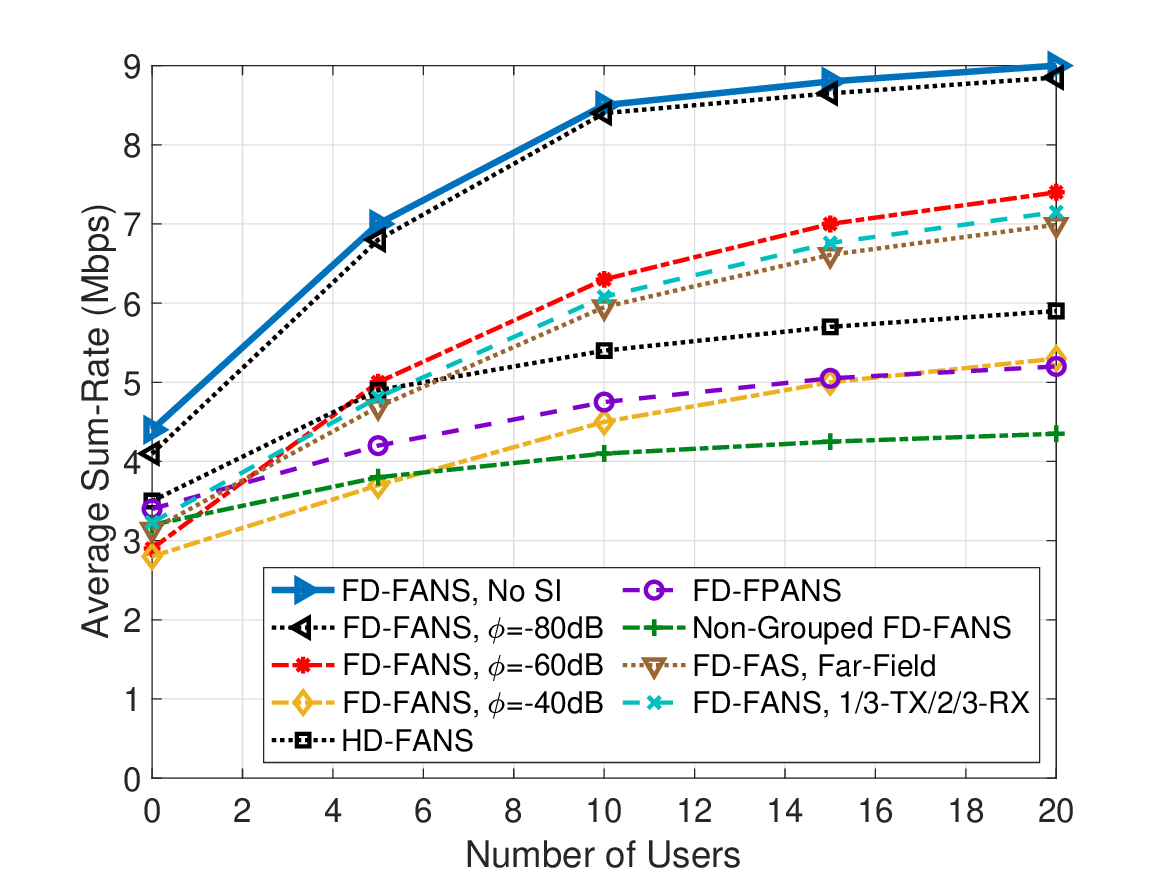}
    \caption{ASR versus $K$ with different values of $\phi$, $P_\mathrm{peak} = 2P_\mathrm{avg}$, and $P_\mathrm{avg} = 20$ dBm}
    \label{f3}
        \vspace{-1em}
\end{figure}

Fig. \ref{f3} presents ASR versus the number of users $K$ for different residual SI levels $\phi$, with $P_{\mathrm{avg}}=20$ dBm and $P_{\mathrm{peak}}=2P_{\mathrm{avg}}$. For fixed $P_{\mathrm{avg}}$ and $P_{\mathrm{peak}}$, the ASR grows nearly linearly with $K$ before tapering, reflecting gains from improved energy-harvesting efficiency and multiuser diversity tempered by slot and aperture sharing. The curves for different $\phi$ remain approximately parallel, indicating that the proposed method maintains SI control as the system scales. The advantage of FD-FANS over HD, FD-FPANS, and the non-grouped FD baseline widens with $K$, confirming that mobility plus grouping unlock additional spatial DoFs for concurrent WET and uplink return. Compared with the far-field counterpart, the proposed FD-FANS achieves about \(0.35\) Mbps higher ASR at \(K=10\) and about \(0.4\) Mbps at \(K=20\). It also exceeds the baseline with asymmetric partitions by roughly \(0.2\) Mbps at \(K=10\) and about \(0.25\) Mbps at \(K=20\).

\begin{figure}[t]
    \centering
    \includegraphics[width=90mm]{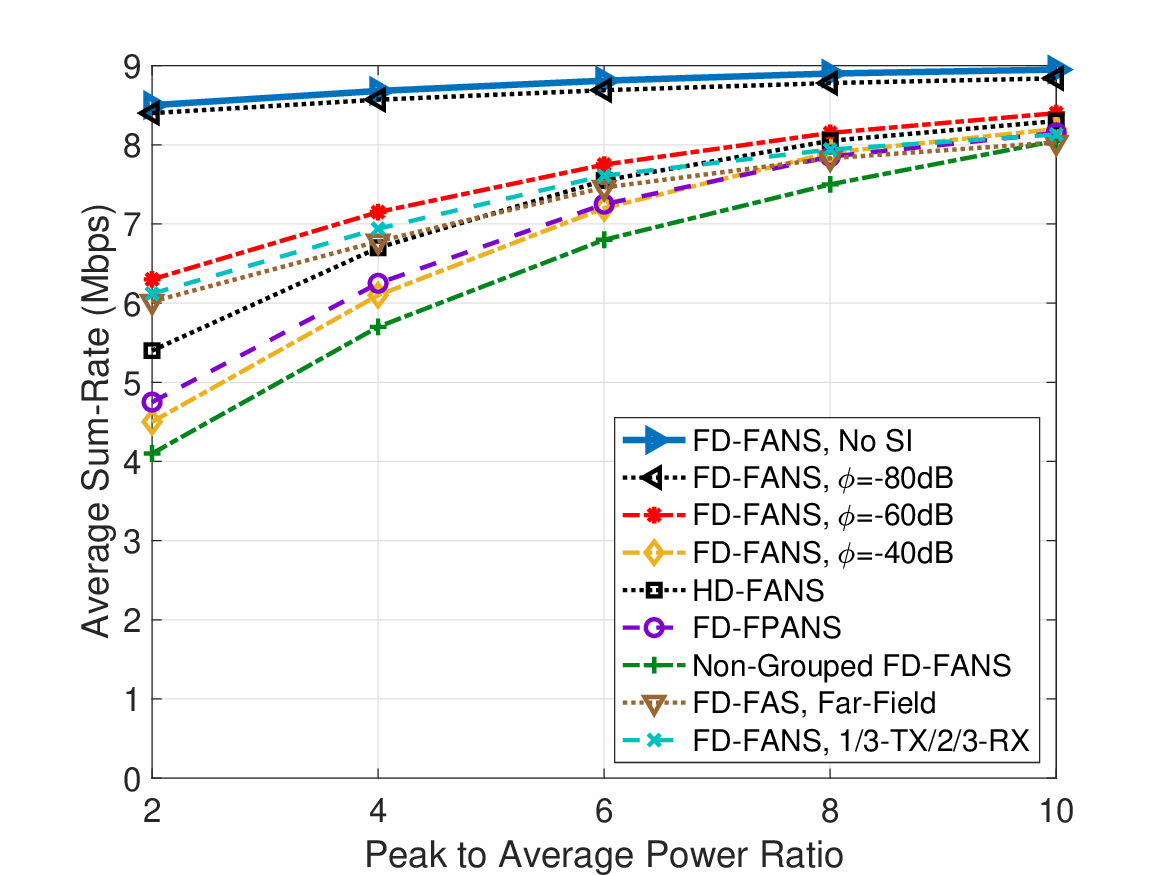}
    \caption{ASR versus $P_\mathrm{peak}/P_\mathrm{avg}$ with different values of $\phi$, $P_\mathrm{avg} = 20$ dBm, and $K = 10$}
    \label{f4}
    \vspace{-1em}
\end{figure}

Fig. \ref{f4} shows ASR versus the PAPR $P_{\mathrm{peak}}/P_{\mathrm{avg}}$ for different residual SI levels $\phi$, with $K=10$ and $P_{\mathrm{avg}}=20$ dBm. Increasing $P_{\mathrm{peak}}/P_{\mathrm{avg}}$ yields clear gains, especially at moderate SI, since higher peaks allow tighter concentration of downlink WET in fewer slots, boosting harvested energy without proportionally extending the vulnerable uplink window. As the ratio grows, the marginal gain flattens due to residual SI and duty-cycle limits. Across all settings, FD-FANS achieves the best performance. Compared with the far-field baseline, the proposed FD-FANS attains about \(0.34\) Mbps more ASR at \(P_{\mathrm{peak}}/P_{\mathrm{avg}}=6\) and about \(0.22\) Mbps at \(P_{\mathrm{peak}}/P_{\mathrm{avg}}=10\). It also remains slightly better than the baseline with asymmetric partitions, with gains of roughly \(0.2\) Mbps at \(P_{\mathrm{peak}}/P_{\mathrm{avg}}=6\) and about \(0.1\) Mbps at \(P_{\mathrm{peak}}/P_{\mathrm{avg}}=10\). By contrast, HD-FANS and FD-FPANS accrue only modest improvements (typically $<0.5$ Mbps), while the proposed system still translates peak capability into throughput more effectively than the baselines.

\begin{figure}[t]
    \centering
    \includegraphics[width=90mm]{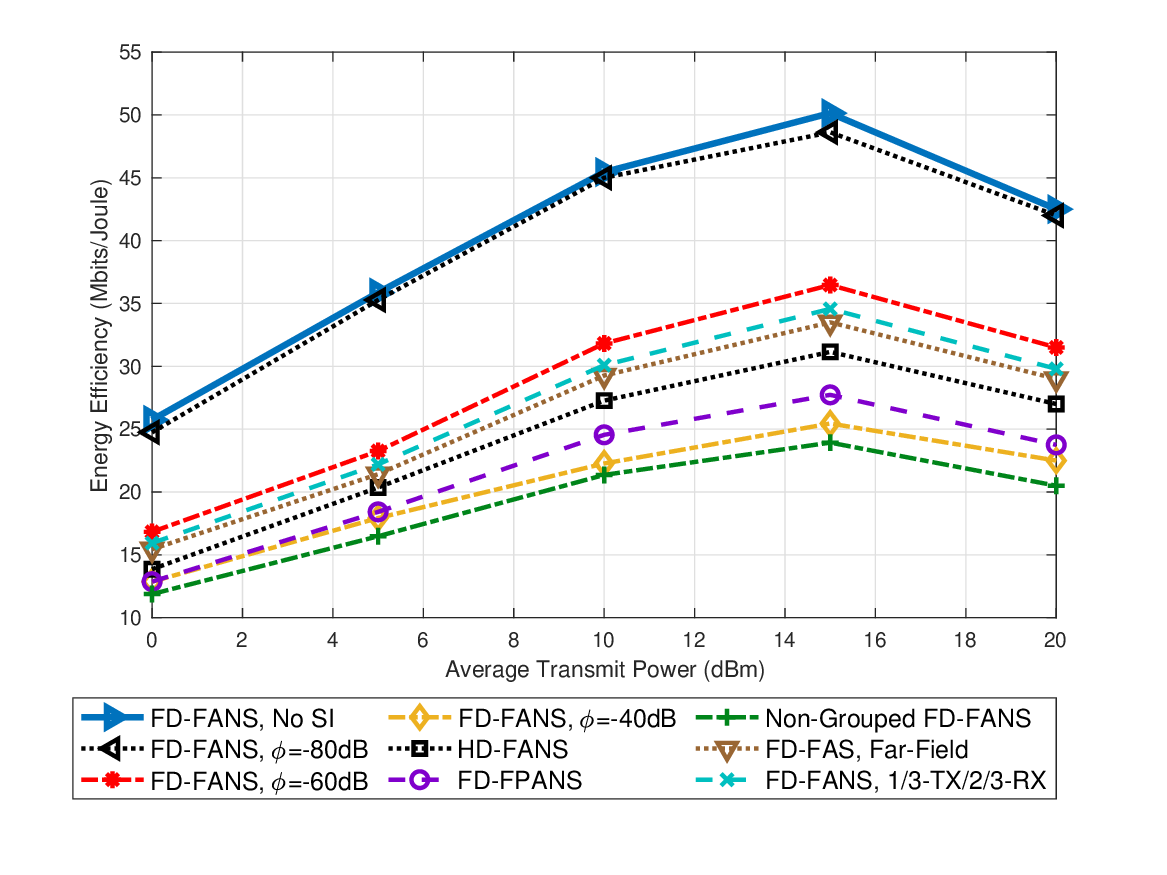}
    \caption{EE versus \(P_{\mathrm{avg}}\) with different values of \(\phi\), \(P_{\mathrm{peak}}=2P_{\mathrm{avg}}\), and \(K=10\)}
    \label{f6}
            \vspace{-1em}
\end{figure}

Moreover, to provide a fair and consistent comparison of the rate delivered per unit energy consumed across different benchmark schemes, we additionally consider the EE (Mbits/Joule). In the simulations, the EE is defined as $\mathrm{EE} = \frac{R_{\mathrm{sum}}}{P_{\mathrm{tot}}}$, where \(R_{\mathrm{sum}}\) denotes the ASR and $P_{\mathrm{tot}} = P_{\mathrm{avg}} + P_c$ is the total power consumption, with \(P_c\) denoting the constant circuit power. We set \(P_c = 0.1~\mathrm{W}\). Fig. \ref{f6} shows the EE versus \(P_{\mathrm{avg}}\) under the same setting as Fig. \ref{f2}. It can be observed that the EE of all schemes first increases and then decreases, which indicates that beyond a certain transmit-power level, the additional ASR gain can no longer compensate for the increased power consumption. For example, the proposed FD-FANS with \(\phi=-60\) dB improves from about \(16.8\) Mbits/J at \(P_{\mathrm{avg}}=0\) dBm to about \(36.5\) Mbits/J at \(15\) dBm, and then decreases to about \(31.5\) Mbits/J at \(20\) dBm. Among all compared schemes, the proposed FD-FANS consistently achieves the highest EE. These results indicate that the ASR improvement of the proposed scheme is not obtained merely by consuming more power, but also by utilizing the available energy more efficiently.

\begin{figure}[t]
    \centering
    \includegraphics[width=90mm]{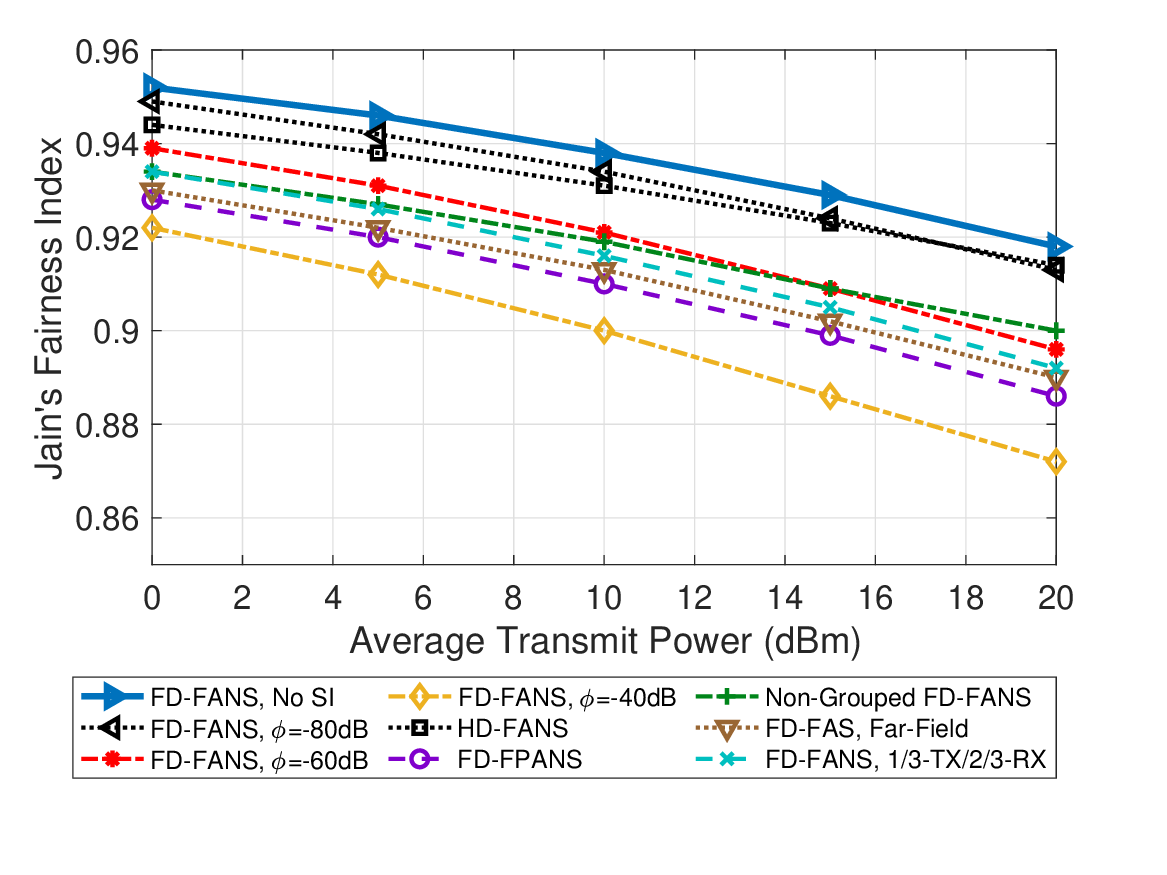}
    \caption{Jain's fairness index versus \(P_{\mathrm{avg}}\) with different values of \(\phi\), \(P_{\mathrm{peak}}=2P_{\mathrm{avg}}\), and \(K=10\).}
    \label{f7}
    \vspace{-1em}
\end{figure}

Finally, to evaluate per-user fairness, we adopt Jain's fairness index, defined as $J = \frac{\left(\sum_{k=1}^{K} R_k\right)^2}{K\sum_{k=1}^{K} R_k^2}$, where \(R_k\) denotes the achievable rate of user \(k\). For each simulation point, the Jain's index is computed from the individual user rates obtained after convergence of the proposed AO algorithm. Fig. \ref{f7} illustrates the Jain's index versus \(P_{\mathrm{avg}}\) with different values of \(\phi\), \(P_{\mathrm{peak}}=2P_{\mathrm{avg}}\), and \(K=10\). It is observed that the Jain's index decreases gradually with \(P_{\mathrm{avg}}\) for all schemes, indicating that the sum-rate-oriented design tends to allocate more resources to favorable users in the high-power regime. Nevertheless, all compared schemes still maintain relatively high fairness values, mostly above \(0.88\), which shows that the ASR gain is not obtained at the cost of severe user-rate imbalance. For example, at \(P_{\mathrm{avg}}=15\) dBm, the proposed FD-FANS with \(\phi=-60\) dB achieves a fairness index of about \(0.909\), compared with about \(0.899\) for FD-FPANS. In addition, the proposed scheme remains close to HD-FANS and non-grouped FD-FANS in fairness, indicating that its ASR and EE improvements are achieved while still preserving a satisfactory user-rate balance.

\section{Conclusion} \label{s6}

In this paper, we proposed an AO-based joint resource allocation framework for the FD-FANS. By jointly considering spherical-wave NF propagation, WET, residual SI, antenna mobility, and TX/RX grouping, we formulated a WSR maximization problem over time--power allocation, antenna positions, and binary group selection under practical geometric and power constraints. To tackle the resulting non-convex mixed-integer problem, an efficient AO framework based on MM/SCA was developed, which achieves monotonic objective improvement and converges to a stationary solution of the relaxed problem. Numerical results demonstrated that the proposed FD-FANS consistently outperforms several benchmark schemes in terms of ASR, EE, and user fairness, while also exhibiting robustness under residual SI and imperfect CSI. Future work may extend this framework by considering geometry-dependent SI coupling, more general multiple-access schemes, and broader NF communication scenarios with frequency-selective channels, hardware impairments, or ISAC functionalities.


\appendix
\section*{Derivation of the Closed-Form Updates in (36)--(39)}\label{ap1}

For the perfect-SIC case ($\tau=0$), problem \eqref{e31} is convex, and its Lagrangian is given by \eqref{e32}. 
By grouping terms associated with each pair $(\delta_k,E_k)$, we rewrite it as
\begin{equation}
L(\boldsymbol{\delta},\mathbf{E},\Lambda,\nu)
=
\sum_{k=0}^{K} L_k(\delta_k,E_k,\Lambda,\nu)
+\Lambda-\nu P_{\mathrm{avg}},
\label{eq:app_short_1}
\end{equation}
where
\begin{equation}
L_k(\delta_k,E_k,\Lambda,\nu)
=
\begin{cases}
-\Lambda \delta_0+\nu E_0,&k=0,\\[1mm]
\alpha_k \delta_k \log_2 \!\left(
1+\dfrac{\omega_k}{\delta_k}(P_{\mathrm{avg}}-E_k)
\right)\\ \quad \quad \quad -\Lambda \delta_k-\nu E_k,&k\in\mathcal{K}.
\end{cases}
\label{eq:app_short_2}
\end{equation}
Hence, for given dual variables $(\Lambda,\nu)$, the dual function can be obtained by maximizing each $L_k$ separately subject to
\begin{equation}
0\le \delta_k \le 1,\qquad 0\le E_k \le P_{\mathrm{peak}}\delta_k.
\label{eq:app_short_3}
\end{equation}

For $k\in\mathcal{K}$, fixing $\delta_k$ and setting $\partial L_k/\partial E_k=0$ yields
\begin{equation}
\frac{\alpha_k\omega_k}
{1+\frac{\omega_k}{\delta_k}(P_{\mathrm{avg}}-E_k)}
=
\nu \ln 2,
\label{eq:app_short_4}
\end{equation}
from which the optimal energy allocation is
\begin{equation}
\hat{E}_k
=
\min\!\left[
\left(
P_{\mathrm{avg}}+\frac{\delta_k}{\omega_k}
-\frac{\alpha_k\delta_k}{\nu \ln 2}
\right)^+,
\,
P_{\mathrm{peak}}\delta_k
\right],\quad k\in\mathcal{K}.
\label{eq:app_short_5}
\end{equation}

Next, fixing $E_k$ and setting $\partial L_k/\partial \delta_k=0$, define
\begin{equation}
q_k \triangleq \frac{\omega_k}{\delta_k}(P_{\mathrm{avg}}-E_k),
\label{eq:app_short_6}
\end{equation}
which leads to
\begin{equation}
f(q_k)\triangleq \ln(1+q_k)-\frac{q_k}{1+q_k}
=
\frac{\Lambda \ln 2}{\alpha_k}.
\label{eq:app_short_7}
\end{equation}
Let $q_k^\star$ denote the solution of \eqref{eq:app_short_7}. Then, the optimal time allocation is
\begin{equation}
\hat{\delta}_k
=
\min\!\left[
\left(
\frac{\omega_k}{q_k^\star}(P_{\mathrm{avg}}-\hat{E}_k)
\right)^+,
\,1
\right],\quad k\in\mathcal{K}.
\label{eq:app_short_8}
\end{equation}

For $k=0$, we have
\begin{equation}
L_0(\delta_0,E_0,\Lambda,\nu)=-\Lambda \delta_0+\nu E_0.
\label{eq:app_short_9}
\end{equation}
Since $0\le E_0\le P_{\mathrm{peak}}\delta_0$, the optimal $E_0$ is
\begin{equation}
\hat{E}_0=
\begin{cases}
P_{\mathrm{peak}}\hat{\delta}_0, & \nu>0,\\
0, & \text{otherwise},
\end{cases}
\label{eq:app_short_10}
\end{equation}
and the optimal $\delta_0$ is
\begin{equation}
\hat{\delta}_0=
\begin{cases}
1, & \nu>0\ \text{and}\ (-\Lambda+\nu P_{\mathrm{peak}})>0,\\
0, & \text{otherwise}.
\end{cases}
\label{eq:app_short_11}
\end{equation}

Combining \eqref{eq:app_short_5}, \eqref{eq:app_short_8}, \eqref{eq:app_short_10}, and \eqref{eq:app_short_11}, we obtain the closed-form updates in \eqref{eq:tau0}--\eqref{eq:Ei}.

\bibliographystyle{ieeetr}
\bibliography{ref}

@inproceedings{fang2021secure,
  title={Secure intelligent reflecting surface assisted UAV communication networks},
  author={Fang, Junhao and Yang, Zhaohui and Anjum, Nasreen and Hu, Ye and Asgari, Hamid and Shikh-Bahaei, Mohammad},
  booktitle={2021 IEEE International Conference on Communications Workshops (ICC Workshops)},
  pages={1--6},
  year={2021},
  organization={IEEE}
}

@article{wu2025fluid,
  title={Fluid antenna systems enabling {6G}: Principles, applications, and research directions},
  author={Wu, Tuo and Zhi, Kangda and Yao, Junteng and Lai, Xiazhi and Zheng, Jianchao and Niu, Hong and Elkashlan, Maged and Wong, Kai-Kit and Chae, Chan-Byoung and Ding, Zhiguo and others},
  journal={IEEE Wirel. Commun.},
  year={Dec. 2025},
  publisher={IEEE},
  doi = {10.1109/MWC.2025.3629597},
  note = {Early Access}
}

@article{wu2025fas,
  title={{FAS}-{RIS} for {V2X}: Unlocking Realistic Performance Analysis with Finite Elements},
  author={Wu, Tuo and Lai, Xiazhi and Zhi, Kangda and Elkashlan, Maged and Al-Dhahir, Naofal and Valenti, Matthew C and Adachi, Fumiyuki},
  journal={IEEE Trans. Veh. Technol.},
  year={Dec. 2025},
  publisher={IEEE},
  doi = {10.1109/TVT.2025.3647789},
  note = {Early Access}
}

@article{wu2025variable,
  title={Variable Block-Correlation Modeling and Optimization for Secrecy Analysis in Fluid Antenna Systems},
  author={Wu, Tuo and Luk, Kwai-Man and Tang, Jie and Wong, Kai-Kit and Zheng, Jianchao and Liu, Baiyang and Morales-Jimenez, David and Elkashlan, Maged and Tong, Kin-Fai and Chae, Chan-Byoung and others},
  journal={arXiv preprint},
  year= 2025,
  archivePrefix = {arXiv},
  pages = {arXiv:2510.03594}
}

@article{wu2025scalable,
  title={Scalable {FAS}: A new paradigm for array signal processing},
  author={Wu, Tuo and Tian, Ye and Tang, Jie and Zhi, Kangda and Elkashlan, Maged and Tong, Kin-Fai and Al-Dhahir, Naofal and Chae, Chan-Byoung and Valenti, Matthew C and Karagiannidis, George K and others},
  journal={arXiv preprint},
  year= 2025,
  archivePrefix = {arXiv},
  pages = {arXiv:2508.10831}
}

@article{zhang2019neural,
  title={A neural network prediction-based adaptive mode selection scheme in full-duplex cognitive networks},
  author={Zhang, Yirun and Hou, Jiancao and Towhidlou, Vahid and Shikh-Bahaei, Mohammad R},
  journal={IEEE Transactions on Cognitive Communications and Networking},
  volume={5},
  number={3},
  pages={540--553},
  year={2019},
  publisher={IEEE}
}

@article{olfat2008optimum,
  title={Optimum power and rate adaptation for MQAM in Rayleigh flat fading with imperfect channel estimation},
  author={Olfat, Ali and Shikh-Bahaei, Mohammad},
  journal={IEEE transactions on vehicular technology},
  volume={57},
  number={4},
  pages={2622--2627},
  year={2008},
  publisher={IEEE}
}

@article{bobarshad2010low,
  title={A low-complexity analytical modeling for cross-layer adaptive error protection in video over WLAN},
  author={Bobarshad, Hossein and van der Schaar, Mihaela and Shikh-Bahaei, Mohammad R},
  journal={IEEE Transactions on Multimedia},
  volume={12},
  number={5},
  pages={427--438},
  year={2010},
  publisher={IEEE}
}

@article{nehra2010spectral,
  title={Spectral efficiency of adaptive MQAM/OFDM systems with CFO over fading channels},
  author={Nehra, Krishna and Shikh-Bahaei, Mohammad},
  journal={IEEE transactions on vehicular technology},
  volume={60},
  number={3},
  pages={1240--1247},
  year={2010},
  publisher={IEEE}
}

@inproceedings{bobarshad2009m,
  title={M/M/1 queuing model for adaptive cross-layer error protection in WLANs},
  author={Bobarshad, Hossein and Shikh-Bahaei, Mohammad},
  booktitle={2009 IEEE Wireless Communications and Networking Conference},
  pages={1--6},
  year={2009},
  organization={IEEE}
}

@article{shikh2007joint,
  title={Joint optimization Of “Transmission Rate” and “Outer-Loop SNR Target” adaptation over fading channels},
  author={Shikh-Bahaei, Mohammad},
  journal={IEEE Transactions on Communications},
  volume={55},
  number={3},
  pages={398--403},
  year={2007},
  publisher={IEEE}
}

@inproceedings{nehra2010cross,
  title={Cross-layer design for interference-limited spectrum sharing systems},
  author={Nehra, Krishna and Shadmand, Amir and Shikh-Bahaei, Mohammad},
  booktitle={2010 IEEE Global Telecommunications Conference GLOBECOM 2010},
  pages={1--5},
  year={2010},
  organization={IEEE}
}

@inproceedings{shadmand2010multi,
  title={Multi-user time-frequency downlink scheduling and resource allocation for LTE cellular systems},
  author={Shadmand, Amir and Shikh-Bahaei, Mohammad},
  booktitle={2010 IEEE Wireless Communication and Networking Conference},
  pages={1--6},
  year={2010},
  organization={IEEE}
}

@inproceedings{kobravi2007cross,
  title={Cross-layer adaptive ARQ and modulation tradeoffs},
  author={Kobravi, Alireza and Shikh-Bahaei, Mohammd},
  booktitle={2007 IEEE 18th International Symposium on Personal, Indoor and Mobile Radio Communications},
  pages={1--5},
  year={2007},
  organization={IEEE}
}

@inproceedings{olfat2005optimum,
  title={Optimum power and rate adaptation with imperfect channel estimation for MQAM in rayleigh flat fading channel},
  author={Olfat, Ali and Shikh-Bahaei, Mohammad},
  booktitle={VTC-2005-Fall. 2005 IEEE 62nd Vehicular Technology Conference, 2005.},
  volume={4},
  pages={2468--2471},
  year={2005},
  organization={IEEE}
}

@article{towhidlou2017improved,
  title={Improved cognitive networking through full duplex cooperative ARQ and HARQ},
  author={Towhidlou, Vahid and Shikh-Bahaei, Mohammad},
  journal={IEEE Wireless Communications Letters},
  volume={7},
  number={2},
  pages={218--221},
  year={2017},
  publisher={IEEE}
}

@article{jia2020channel,
  title={Channel assignment in uplink wireless communication using machine learning approach},
  author={Jia, Guangyu and Yang, Zhaohui and Lam, Hak-Keung and Shi, Jianfeng and Shikh-Bahaei, Mohammad},
  journal={IEEE Communications Letters},
  volume={24},
  number={4},
  pages={787--791},
  year={2020},
  publisher={IEEE}
}

@ARTICLE{2024arXiv241202538Y,
       author = {{Yang}, Zhaohui and {Xu}, Wei and {Liang}, Le and {Cui}, Yuanhao and {Qin}, Zhijin and {Debbah}, Merouane},
        title = "{On Privacy, Security, and Trustworthiness in Distributed Wireless Large {AI} Models (WLAM)}",
      journal = {arXiv e-prints},
         year = 2024,
          pages = {arXiv:2412.02538},
          doi = {10.48550/arXiv.2412.02538},
archivePrefix = {arXiv},
       eprint = {2412.02538},
 primaryClass = {cs.IT}
}

@article{yao2023robust,
  title={Robust beamforming design for {RIS}-aided cell-free systems with {CSI} uncertainties and capacity-limited backhaul},
  author={Yao, Jiacheng and Xu, Jindan and Xu, Wei and Ng, Derrick Wing Kwan and Yuen, Chau and You, Xiaohu},
  journal={IEEE Trans. Commun.},
  volume={71},
  number={8},
  pages={4636--4649},
  year={May 2023},
  publisher={IEEE}
}

@article{du2025full,
  title={A full-duplex based integrated sensing and communication survey: Principles, key techniques, and receiver design},
  author={Du, Changhao and Zhang, Hongru and Zhang, Xueting and Zhao, Zizheng and Yang, Jie and Zhang, Xinyuan and Xing, Zhifang and Feng, Zhipeng and Zuo, Shiyu and Xu, Canni and others},
  journal={IEEE Commun. Surv. Tuts.},
  year={Jun. 2025},
  publisher={IEEE}
}

@inproceedings{nazar2025full,
  title={Full-Duplex Beamforming Optimization for Near-Field {ISAC}},
  author={Nazar, Ahsan and Shaikhanov, Zhambyl and Ulukus, Sennur},
  booktitle={Proc. 2025 IEEE Mil. Commun. Conf. (MILCOM)},
  pages={687--692},
  year={2025},
  organization={IEEE}
}

@inproceedings{sheemar2025near,
  title={Near-Field Full Duplex {XL} {MIMO} with Reconfigurable Holographic Surfaces},
  author={Sheemar, Chandan Kumar and Khan, Wali Ullah and Solanki, Sourabh and Alexandropoulos, George C and Abdullah, Zaid and Chatzinotas, Symeon},
  booktitle={Proc. IEEE 36th Int. Symp. Pers., Indoor Mobile Radio Commun. (PIMRC)},
  pages={1--6},
  year={2025},
  organization={IEEE}
}

@article{shao20256d,
  title={{6D} movable antenna enhanced wireless network via discrete position and rotation optimization},
  author={Shao, Xiaodan and Zhang, Rui and Jiang, Qijun and Schober, Robert},
  journal={IEEE J. Sel. Areas Commun.},
  volume={43},
  number={3},
  pages={674--687},
  year={Jan. 2025},
  publisher={IEEE}
}

@article{liu2023near,
  title={Near-field communications: {A} tutorial review},
  author={Liu, Yuanwei and Wang, Zhaolin and Xu, Jiaqi and Ouyang, Chongjun and Mu, Xidong and Schober, Robert},
  journal={IEEE Open J. Commun. Soc.},
  volume={4},
  pages={1999--2049},
  year={Aug. 2023},
  publisher={IEEE}
}

@article{yao2024wireless,
  title={Wireless federated learning over resource-constrained networks: Digital versus analog transmissions},
  author={Yao, Jiacheng and Xu, Wei and Yang, Zhaohui and You, Xiaohu and Bennis, Mehdi and Poor, H Vincent},
  journal={IEEE Trans. Wireless Commun.},
  volume={23},
  number={10},
  pages={14020--14036},
  year={Jun. 2024},
  publisher={IEEE}
}

@article{ju2014optimal1,
  title={Optimal Resource Allocation in Full-Duplex Wireless-Powered Communication Network},
  author={Ju, Hyungsik and Zhang, Rui},
  journal={arXiv preprint arXiv:1403.2580},
  year={2014}
}

@article{orhan2014energy,
  title={Energy harvesting broadband communication systems with processing energy cost},
  author={Orhan, Oner and G{\"u}nd{\"u}z, Deniz and Erkip, Elza},
  journal={IEEE Trans. Wireless Commun.},
  volume={13},
  number={11},
  pages={6095--6107},
  year={Nov. 2014},
  publisher={IEEE}
}

@article{chen2024joint,
  title={Joint beamforming and antenna design for near-field fluid antenna system},
  author={Chen, Yixuan and Chen, Mingzhe and Xu, Hao and Yang, Zhaohui and Wong, Kai-Kit and Zhang, Zhaoyang},
  journal={IEEE Wireless Commun. Lett.},
  volume={14},
  number={2},
  pages={415--419},
  year={Nov. 2024},
  publisher={IEEE}
}

@article{towhidlou2018adaptive,
  title={Adaptive full-duplex communications in cognitive radio networks},
  author={Towhidlou, Vahid and Shikh-Bahaei, Mohammad},
  journal={IEEE Trans. Veh. Technol.},
  volume={67},
  number={9},
  pages={8386--8395},
  year={Jun. 2018},
  publisher={IEEE}
}

@article{zhang2020ensemble,
  title={On ensemble learning-based secure fusion strategy for robust cooperative sensing in full-duplex cognitive radio networks},
  author={Zhang, Yirun and Wu, Qirui and Shikh-Bahaei, Mohammad R},
  journal={IEEE Trans. Commun.},
  volume={68},
  number={10},
  pages={6086--6100},
  year={Jun. 2020},
  publisher={IEEE}
}

@article{zhang2013mimo,
  title={{MIMO} broadcasting for simultaneous wireless information and power transfer},
  author={Zhang, Rui and Ho, Chin Keong},
  journal={IEEE Trans. Wireless Commun.},
  volume={12},
  number={5},
  pages={1989--2001},
  year={Mar. 2013},
  publisher={IEEE}
}

@article{nasir2013relaying,
  title={Relaying protocols for wireless energy harvesting and information processing},
  author={Nasir, Ali A and Zhou, Xiangyun and Durrani, Salman and Kennedy, Rodney A},
  journal={IEEE Trans. Wireless Commun.},
  volume={12},
  number={7},
  pages={3622--3636},
  year={Jul. 2013},
  publisher={IEEE}
}

@article{an2024exploiting,
  title={Exploiting multi-layer refracting {RIS}-assisted receiver for {HAP}-{SWIPT} networks},
  author={An, Kang and Sun, Yifu and Lin, Zhi and Zhu, Yonggang and Ni, Wanli and Al-Dhahir, Naofal and Wong, Kai-Kit and Niyato, Dusit},
  journal={IEEE Trans. Wireless Commun.},
  volume={23},
  number={10},
  pages={12638--12657},
  year={May 2024},
  publisher={IEEE}
}

@article{zhang2023energy,
  title={Energy efficiency maximization in {RIS}-assisted {SWIPT} networks with {RSMA}: A {PPO}-based approach},
  author={Zhang, Ruichen and Xiong, Ke and Lu, Yang and Fan, Pingyi and Ng, Derrick Wing Kwan and Letaief, Khaled B},
  journal={IEEE J. Sel. Areas Commun.},
  volume={41},
  number={5},
  pages={1413--1430},
  year={Jan. 2023},
  publisher={IEEE}
}

@inproceedings{zhou2024location,
  title={Location Optimization for Fluid Antenna-Assisted Near-Field System},
  author={Zhou, Jingxuan and Yang, Yinchao and Yang, Zhaohui and Xu, Wei and Shikh-Bahaei, Mohammad},
  booktitle={Proc. IEEE Glob. Commun. Workshops (GC WKSHPS)},
  pages={1--6},
  year={2024},
  organization={IEEE}
}

@inproceedings{yang2025enhanced,
  title={Enhanced Metaverse Experiences: Fluid Antenna-Enabled Integrated Sensing, Computing, and Semantic Communication},
  author={Yang, Yinchao and Zhou, Jingxuan and Yang, Zhaohui and Shikh-Bahaei, Mohammad},
  booktitle={Proc. IEEE Conf. Comput. Commun. Workshops(INFOCOM WKSHPS)},
  pages={1--6},
  year={2025},
  organization={IEEE}
}

@article{chen20235g,
  title={{5G}-advanced toward {6G}: Past, present, and future},
  author={Chen, Wanshi and Lin, Xingqin and Lee, Juho and Toskala, Antti and Sun, Shu and Chiasserini, Carla Fabiana and Liu, Lingjia},
  journal={IEEE J. Sel. Areas Commun.},
  volume={41},
  number={6},
  pages={1592--1619},
  year={Jun. 2023},
  publisher={IEEE}
}

@article{li2019full,
  title={Full-duplex cooperative {NOMA} relaying systems with {I/Q} imbalance and imperfect {SIC}},
  author={Li, Xingwang and Liu, Meng and Deng, Chao and Mathiopoulos, P Takis and Ding, Zhiguo and Liu, Yuanwei},
  journal={IEEE Wireless Commun. Lett.},
  volume={9},
  number={1},
  pages={17--20},
  year={Sep. 2019},
  publisher={IEEE}
}

@article{ju2014optimal,
  title={Optimal resource allocation in full-duplex wireless-powered communication network},
  author={Ju, Hyungsik and Zhang, Rui},
  journal={IEEE Trans. Commun.},
  volume={62},
  number={10},
  pages={3528--3540},
  year={Sep. 2014},
  publisher={IEEE}
}

@article{sabharwal2014band,
  title={In-band full-duplex wireless: Challenges and opportunities},
  author={Sabharwal, Ashutosh and Schniter, Philip and Guo, Dongning and Bliss, Daniel W and Rangarajan, Sampath and Wichman, Risto},
  journal={IEEE J. Sel. Areas Commun.},
  volume={32},
  number={9},
  pages={1637--1652},
  year={Jun. 2014},
  publisher={IEEE}
}

@article{shao20246d,
  title={{6D} movable antenna based on user distribution: Modeling and optimization},
  author={Shao, Xiaodan and Jiang, Qijun and Zhang, Rui},
  journal={IEEE Trans. Wireless Commun.},
  year={Nov. 2024},
  publisher={IEEE}
}

@article{qin2024antenna,
  title={Antenna positioning and beamforming design for fluid antenna-assisted multi-user downlink communications},
  author={Qin, Haoran and Chen, Wen and Li, Zhendong and Wu, Qingqing and Cheng, Nan and Chen, Fangjiong},
  journal={IEEE Wireless Commun. Lett.},
  volume={13},
  number={4},
  pages={1073--1077},
  year={Jan. 2024},
  publisher={IEEE}
}

@article{new2024tutorial,
  title={A tutorial on fluid antenna system for {6G} networks: Encompassing communication theory, optimization methods and hardware designs},
  author={New, Wee Kiat and Wong, Kai-Kit and Xu, Hao and Wang, Chao and Ghadi, Farshad Rostami and Zhang, Jichen and Rao, Junhui and Murch, Ross and Ram{\'\i}rez-Espinosa, Pablo and Morales-Jimenez, David and others},
  journal={IEEE Commun. Surv. Tuts.},
  volume={27},
  number={4},
  pages={2325 - 2377},
  year={Nov. 2024},
  publisher={IEEE}
}

@article{zhu2023modeling,
  title={Modeling and performance analysis for movable antenna enabled wireless communications},
  author={Zhu, Lipeng and Ma, Wenyan and Zhang, Rui},
  journal={IEEE Trans. Wireless Commun.},
  volume={23},
  number={6},
  pages={6234--6250},
  year={Nov. 2023},
  publisher={IEEE}
}

@article{wong2020fluid,
  title={Fluid antenna systems},
  author={Wong, Kai-Kit and Shojaeifard, Arman and Tong, Kin-Fai and Zhang, Yangyang},
  journal={IEEE Trans. Wireless Commun.},
  volume={20},
  number={3},
  pages={1950--1962},
  year={Nov. 2020},
  publisher={IEEE}
}

@article{ding2024utilizing,
  title={Utilizing imperfect resolution of near-field beamforming: A hybrid-{NOMA} perspective},
  author={Ding, Zhiguo and Poor, H Vincent},
  journal={IEEE Commun. Lett.},
  volume={28},
  number={7},
  pages={1718--1722},
  year={May 2024},
  publisher={IEEE}
}

@article{zhou2024near,
  title={Near-field extremely large-scale {STAR-RIS} enabled integrated sensing and communications},
  author={Zhou, Jingxuan and Yang, Yinchao and Yang, Zhaohui and Shikh-Bahaei, Mohammad Reza},
  journal={IEEE Trans. Green Commun. Netw.},
  volume={9},
  number={1},
  pages={404--416},
  year={Sep. 2024},
  publisher={IEEE}
}

@article{mou2022near,
  title={Near-field wireless power transfer technology for unmanned aerial vehicles: A systematical review},
  author={Mou, Xiaolin and Gladwin, Daniel and Jiang, Jing and Li, Kang and Yang, Zhile},
  journal={IEEE J. Emerg. Sel. Top. Ind. Electron.}, 
  volume={4},
  number={1},
  pages={147--158},
  year={Nov. 2022},
  publisher={IEEE}
}

@article{zhang20236g,
  title={{6G} wireless communications: From far-field beam steering to near-field beam focusing},
  author={Zhang, Haiyang and Shlezinger, Nir and Guidi, Francesco and Dardari, Davide and Eldar, Yonina C},
  journal={IEEE Commun. Mag.},
  volume={61},
  number={4},
  pages={72--77},
  year={Mar. 2023},
  publisher={IEEE}
}

@article{smida2023full,
  title={Full-duplex wireless for {6G}: Progress brings new opportunities and challenges},
  author={Smida, Besma and Sabharwal, Ashutosh and Fodor, G{\'a}bor and Alexandropoulos, George C and Suraweera, Himal A and Chae, Chan-Byoung},
  journal={IEEE J. Sel. Areas Commun.},
  volume={41},
  number={9},
  pages={2729--2750},
  year={Jun. 2023},
  publisher={IEEE}
}

@article{yang2025fluid,
  title={Fluid Antenna-Enabled Near-Field Integrated Sensing, Computing and Semantic Communication for Emerging Applications},
  author={Yang, Yinchao and Zhou, Jingxuan and Yang, Zhaohui and Shikh-Bahaei, Mohammad},
  journal={IEEE Trans. on Cogn. Commun. Netw.},
  year={Jul. 2025},
  publisher={IEEE}
}

@article{wang2023road,
  title={On the road to {6G}: Visions, requirements, key technologies, and testbeds},
  author={Wang, Cheng-Xiang and You, Xiaohu and Gao, Xiqi and Zhu, Xiuming and Li, Zixin and Zhang, Chuan and Wang, Haiming and Huang, Yongming and Chen, Yunfei and Haas, Harald and others},
  journal={IEEE Commun. Surv. Tuts.},
  volume={25},
  number={2},
  pages={905--974},
  year={Feb. 2023},
  publisher={IEEE}
}

@article{wang2024tutorial,
  title={A tutorial on extremely large-scale {MIMO} for {6G}: Fundamentals, signal processing, and applications},
  author={Wang, Zhe and Zhang, Jiayi and Du, Hongyang and Niyato, Dusit and Cui, Shuguang and Ai, Bo and Debbah, M{\'e}rouane and Letaief, Khaled B and Poor, H Vincent},
  journal={IEEE Commun. Surv. Tuts.},
  volume={26},
  number={3},
  pages={1560--1605},
  year={Jan. 2024},
  publisher={IEEE}
}

@article{jiang2024terahertz,
  title={Terahertz communications and sensing for {6G} and beyond: A comprehensive review},
  author={Jiang, Wei and Zhou, Qiuheng and He, Jiguang and Habibi, Mohammad Asif and Melnyk, Sergiy and El-Absi, Mohammed and Han, Bin and Di Renzo, Marco and Schotten, Hans Dieter and Luo, Fa-Long and others},
  journal={IEEE Commun. Surv. Tuts.},
  volume={26},
  number={4},
  pages={2326--2381},
  year={Apr. 2024},
  publisher={IEEE}
}

@article{shi2024ris,
  title={{RIS}-aided cell-free massive {MIMO} systems for {6G}: Fundamentals, system design, and applications},
  author={Shi, Enyu and Zhang, Jiayi and Du, Hongyang and Ai, Bo and Yuen, Chau and Niyato, Dusit and Letaief, Khaled B and Shen, Xuemin},
  journal={Proc. of the IEEE},
  volume={112},
  number={4},
  pages={331--364},
  year={Jun. 2024},
  publisher={IEEE}
}

@article{lu2024tutorial,
  title={A tutorial on near-field {XL-MIMO} communications toward {6G}},
  author={Lu, Haiquan and Zeng, Yong and You, Changsheng and Han, Yu and Zhang, Jiayi and Wang, Zhe and Dong, Zhenjun and Jin, Shi and Wang, Cheng-Xiang and Jiang, Tao and others},
  journal={IEEE Commun. Surv. Tuts.},
  volume={26},
  number={4},
  pages={2213--2257},
  year={Apr. 2024},
  publisher={IEEE}
}

@article{zhu2023movable,
  title={Movable-antenna array enhanced beamforming: Achieving full array gain with null steering},
  author={Zhu, Lipeng and Ma, Wenyan and Zhang, Rui},
  journal={IEEE Commun. Lett.},
  volume={27},
  number={12},
  pages={3340--3344},
  year={Dec. 2023},
  publisher={IEEE}
}

\vspace{12pt}
\color{red}

\end{document}